\documentclass[12pt,a4paper]{article}
\usepackage{amsmath,amsfonts,amsthm}
\allowdisplaybreaks[3]
\numberwithin{equation}{section}

\usepackage[body={16cm,23cm}]{geometry}
\newcommand{\Ds}{{\mathsf D}}

\newcommand{\Qb}{{\mathbb Q}}

\newcommand{\Ts}{{\mathsf T}}

\newcommand{\Tb}{{\mathbb T}}

\newcommand{\Bm}{{\mathfrak B}}
\newcommand{\Fm}{{\mathfrak F}}

\begin{document}
\title{
Boson-Fermion correspondence, QQ-relations and Wronskian 
solutions of the T-system
}
\author{Zengo Tsuboi
\\[8pt]
{\sl
Pacific Quantum Center, Far Eastern Federal University, 
}
\\
{\sl
Sukhanova 8, Vladivostok, 690950, Russia
%
%
%
}
} 
\date{}
\maketitle
\begin{abstract}
It is known that there is a correspondence between  representations of superalgebras and ordinary (non-graded) algebras. 
Keeping in mind  this type of correspondence between the twisted quantum affine superalgebra $U_{q}(gl(2r|1)^{(2)})$ 
and the non-twisted quantum affine algebra $U_{q}(so(2r+1)^{(1)})$,  
we proposed, in the previous paper \cite{T11}, a Wronskian solution of the T-system for  $U_{q}(so(2r+1)^{(1)})$  
as a reduction (folding) of the Wronskian solution for the non-twisted quantum affine superalgebra $U_{q}(gl(2r|1)^{(1)})$. 
In this paper, we elaborate on this solution, and 
give a proof missing in \cite{T11}.
In particular, we explain its connection to the 
 Cherednik-Bazhanov-Reshetikhin (quantum Jacobi-Trudi) type determinant solution known in \cite{KOS95}. 
We also propose Wronskian-type expressions of T-functions (eigenvalues of transfer matrices) labeled by 
non-rectangular Young diagrams, which are quantum affine algebra analogues of the 
Weyl character formula for $so(2r+1)$. 
We show that T-functions for spinorial representations of $U_{q}(so(2r+1)^{(1)})$  
 are related to reductions of T-functions for asymptotic typical  representations of 
$U_{q}(gl(2r|1)^{(1)})$. 
\end{abstract}
Keywords: Baxter Q-function, Boson-Fermion correspondence, QQ-relation, T-system, 
Wronskian solution 
\\[3pt]
Nuclear Physics B 972 (2021) 115563\\
https://doi.org/10.1016/j.nuclphysb.2021.115563
\section{Introduction}
This paper is an expansion of section 3.7 in \cite{T11}.

The T-system is a system of functional relations among  mutually commuting family of 
transfer matrices of quantum integrable systems. As a difference equation, it is a kind of discrete Toda field equation. 
The T-system was proposed for quantum integrable systems associated with quantum affine algebras (or Yangians) 
\cite{KNS93,KS94-2} (see also \cite{KR87,KP92} for $U_{q}(A_{1}^{(1)})$ case) and superalgebras \cite{T97,T98,T99,T99-2}. 
It also appeared in the study of AdS/CFT integrability \cite{GKV09}. 
In the context of the representation theory, the solutions of the T-system correspond to the q-characters \cite{FR99,FM99} 
of  the Kirillov-Reshetikhin modules. They are quantum affine algebra or Yangian analogues of characters of simple Lie algebras 
with a spectral parameter. See \cite{He05} for a general account on this. 

The T-system was solved  in terms of Cherednik-Bazhanov-Reshetikhin (CBR) 
 or quantum Jacobi-Trudi-type determinants (in some cases, pfaffians) 
for $U_{q}(A^{(1)}_{r})$ in \cite{BR90}, for $U_{q}(B_{r}^{(1)})$, $U_{q}(C_{r}^{(1)})$ and $U_{q}(D_{r}^{(1)})$ in \cite{KNH96}, 
 for $U_{q}(gl(r|s)^{(1)})$ (or $U_{q}(sl(r|s)^{(1)})$) in \cite{T97,T98}, and for $Y(osp(r|2s))$
 \footnote{We have only partial results for this case at the moment.}
  in \cite{T99}. 
 The CBR-type determinants in \cite{KNH96} are defined in terms of rather sparse matrices. 
   In contrast, there are also CBR-type determinant solutions with dense matrices 
  for  $U_{q}(B_{r}^{(1)})$ \cite{KOS95} and  $U_{q}(D_{r}^{(1)})$ \cite{TK96}. 
  Moreover, tableau sum expressions of the solutions are available in many cases (cf. \cite{KNS10}). 
  These solutions are expressed in terms of Baxter Q-functions \cite{Bax72}
  \footnote{The Baxter Q-function is an eigenvalue of a Baxter Q-operator. 
  We call them `Q-function' and `Q-operator', respectively.
  We also call an eigenvalue formula of a transfer matrix `T-function'.} through the analytic Bethe ansatz idea \cite{Re83,KS94-1}. 
  The full set of Q-functions are in general not independent and connected each other via functional relations, called QQ-relations. 
  Thus there is a freedom to choose basic Q-functions to express solutions of the T-system. 
  Choosing a convenient set of basic Q-functions can be an important problem in mathematical physics. 
  The most symmetric form of the solutions with respect to the action of the Weyl group would be Wronskian
  \footnote{In this paper, we use the term Wronskian in rather wide sense. In fact, it is a kind of Casoratian, which is a discrete analogue of the Wronskian.} determinant solutions. 

Wronskian solutions of the T-system were constructed relatively early for $U_{q}(A^{(1)}_{r})$ (or its elliptic counterpart) 
\cite{KLWZ96}, $U_{q}(C^{(1)}_{r})$ \cite{KOSY01}, and 
$U_{q}(A^{(2)}_{r})$ \cite{T02}.  In these cases, the difference L-operators, which generate the 
T-functions for the anti-symmetric representations in the auxiliary space, are finite order difference operators. 
Thus the Baxter equations as kernels of these difference L-operators become finite order difference equations. Then Wronskian expressions of 
T-functions follow from Cramer's rule in linear algebra and are expressed in terms of single indexed Q-functions $Q_{\{a\}}$, 
which are solutions of the Baxter equations. 
In contrast, the difference L-operators for other algebras such as 
$U_{q}(B^{(1)}_{r})$ for $r\ge 2$ \cite{KOS95}, $U_{q}(D^{(1)}_{r})$ for $r \ge 4$ \cite{TK96}, $U_{q}(D^{(2)}_{r+1})$ for $r \ge 3$,  
$U_{q}(D^{(3)}_{4})$ \cite{T02}, 
 $U_{q}(gl(r|s)^{(1)})$ (or $U_{q}(sl^{(1)}(r|s))$) for $r,s \ge 1$ \cite{T97,T98}, $Y(osp(r|2s))$ for $r \ge 3$, $s \ge 1$ \cite{T99}
are infinite order difference operators. This was one of the main obstacles to generalize earlier results. 
In \cite{T09}, we managed to overcome this difficulty and proposed Wronskian solutions of the T-system for $U_{q}(gl(M|N)^{(1)})$. 
We had to use double indexed Q-functions $Q_{\{b,f\}}$ in addition to the single indexed Q-functions $Q_{\{b\}}$ and $Q_{\{f\}}$, 
where the bosonic indexes $b \in \{1,2, \dots , M \}$ and the fermionic indexes $f \in \{M+1,M+2, \dots , M+N \}$ 
carry the parity (grading) of the Lie superalgebra $gl(M|N)$. The key functional relations, which connect these three types of Q-functions, 
are (fermionic) QQ-relations. 
They originated from the particle-hole transformations 
\cite{Wo83}
 in statistical mechanics  
and are related \cite{T98} to odd Weyl reflections \cite{DP85,Se85} of the superalgebra $gl(M|N)$.  
The fermionic QQ-relations are studied in relation to B\"{a}cklund transformations in soliton theory \cite{KSZ07}.  
Moreover, all the $2^{M+N}$ Q-functions $Q_{I}$ labeled by $I \subset \{1,2,\dots , M+N \}$ 
are expressed \cite{T09} as Wronskian determinants in terms of these three types of basic Q-functions. 
This is also the case with the T-functions and in particular solutions of the T-system. 
These results were extended to the T-system for $AdS_{5}/CFT_{4}$ integrability \cite{GKLT10} and  for 
 quantum integrable systems related to infinite dimensional representations of 
$gl(r|s)$\cite{T11}. Connection to Grassmannians was elaborated on in \cite{KLV15}. 
Moreover, Wronskian-like solutions (scalar product of `Q-vectors') of the T-system for 
 $U_{q}({\mathfrak g}^{(1)})$ or $Y({\mathfrak g})$ with ${\mathfrak g}=D_{r}, E_{n}$ ($n=6,7,8$) were 
proposed recently in \cite{ESV20}, and in particular for ${\mathfrak g}=D_{r}$ case,  a  Wronskian-type expression from pure spinors, 
which uses a set of basic Q-functions different from the one used in the solution in \cite{TK96},  
 was presented (see also \cite{FFK20,EV21}). 
 In this paper, we focus on the previously proposed Wronskian determinant solution for $U_{q}(B_{r}^{(1)})$ case \cite{T11},  
 in relation to the CBR-type solution known in \cite{KOS95}. 

It is known \cite{Z97} that there is a correspondence (some kind of Boson-Fermion correspondence) 
between  representations of superalgebras and ordinary (non-graded) algebras. 
Keeping in mind a correspondence between  the twisted quantum affine superalgebra $U_{q}(gl(2r|1)^{(2)})$ and 
the quantum affine algebra $U_{q}(B^{(1)}_{r})$, 
we proposed \cite{T11} a Wronskian solution of the T-system for $U_{q}(B^{(1)}_{r})$ 
as a reduction (folding) of the Wronskian solution for $U_{q}(gl(2r|1)^{(1)})$ \cite{T09}. 
In this paper, we give a proof of the solution, which is missing in \cite{T09}. 
In particular, we will clarify a connection between this Wronskian solution and 
the CBR-type solution known in \cite{KOS95}. 
The above reduction procedures for the tensor representations
\footnote{Representation theoretical background can be found in \cite{BM14}.}
 are straightforward, but there is a subtlety in spinorial representations. 
In fact, we observed  \cite{T11} that the T-functions associated with spinorial representations of $U_{q}(B^{(1)}_{r})$  
are proportional to reductions of T-functions associated with $U_{q}(gl(2r|1)^{(1)})$, which are  
one level
\footnote{On the level of characters, this effectively corresponds to dealing with $gl(2r)$, which is a subalgebra of $gl(2r|1)$. }
 lower than the original T-functions in the sense of the B\"{a}cklund transformations \cite{KSZ07}. 
This means that spinorial representations of $U_{q}(so(2r+1)^{(1)})$ are related to reductions of 
asymptotic representations of $U_{q}(gl(2r|1)^{(1)})$.  

In section 2, we recall a part of \cite{T09,T11}, which deals with T- and Q-functions related to $U_{q}(gl(M|N)^{(1)})$. 
In section 3, we give a Wronskian solution of the T-system for $U_{q}(B^{(1)}_{r})$ as a reduction of the solution given in section 2. 
We show that the QQ-relations for $U_{q}(B^{(1)}_{r})$ follow from the QQ-relations for  $U_{q}(gl(2r|1)^{(1)})$. 
We also consider Wronskian expressions of T-functions labeled by non-rectangular Young diagrams and 
explain their relation to CBR-type formulas associated with $U_{q}(B^{(1)}_{r})$ \cite{KOS95}. 
All these T-functions reduce to Weyl-type character formulas of $U_{q}(B^{(1)}_{r})$ or $Y(B_{r})$ in the limit 
that the Q-functions go to $1$.
Section 4 is devoted to concluding remarks. 

Throughout the paper we assume that the deformation parameter $q$ is not a root of unity when we mention the T-system 
associated with the quantum affine algebra. In addition, 
there is no substantial difference between the Yangian $Y(\mathfrak{g})$ case ($\mathfrak{g}$: a simple Lie (super)algebra) and 
the quantum affine algebra $U_{q}({\mathfrak g}^{(1)})$ case as far as the solutions of the T-system are concerned. 
%
\section{T-and Q-functions for  $U_{q}(gl(M|N)^{(1)})$}
This section is a brief summary of  a part of  \cite{T09,T11}. 
There are  some overlaps in the text with \cite{T09,T11}, with respect to review parts of this paper. 
We will review  miscellaneous formulas on T-and Q-functions 
for quantum integrable models associated with $U_{q}(gl(M|N)^{(1)})$ 
(or $Y(gl(M|N))$). 

\subsection{Notation}
For any $M,N \in \mathbb{Z}_{\ge 0}$, let us introduce sets
\begin{align}
\begin{split}
&{\mathfrak B}=\{1,2,\dots, M \},  \\
& {\mathfrak F}=\{M+1,M+2, \dots, M+N \}, \\
&
{\mathfrak I}={\mathfrak B} \sqcup {\mathfrak F}.
\end{split}
\end{align}
We will use a grading parameter on the set $  {\mathfrak B} \sqcup  {\mathfrak F}$:  
\begin{align} 
p_{a}=1 \quad \text{for} \quad a \in {\mathfrak B}, \qquad 
p_{a}=-1 \quad \text{for} \quad a \in {\mathfrak F}. 
\end{align}
We often use the following notation for subsets of the above sets: 
$B \subset {\mathfrak B}$, 
$F \subset {\mathfrak F}$,  $ I \subset {\mathfrak I}$, and for a matrix:
\begin{align}
(a_{ij})_{i \in B \atop j \in F}=
\begin{pmatrix}
a_{b_{1},f_{1}} & a_{b_{1},f_{2}} & \cdots & a_{b_{1},f_{n}} \\
a_{b_{2},f_{1}} & a_{b_{2},f_{2}} & \cdots & a_{b_{2},f_{n}} \\
\hdotsfor{4} \\
a_{b_{m},f_{1}} & a_{b_{m},f_{2}} & \cdots & a_{b_{m},f_{n}}
\end{pmatrix}
,
\end{align}
where $B=\{b_{1},\dots , b_{m} \}$, $b_{1}<\dots < b_{m}$; 
 $F=\{f_{1},\dots , f_{n} \}$, $f_{1}<\dots < f_{n}$.

Let us consider an arbitrary function $f(u)$ of  $u \in {\mathbb C}$ (the spectral parameter). 
Throughout this paper we use a notation on a shift of the spectral parameter such as  
$f^{[a]}=f(u+a \hbar)$ for an additive shift, and 
$f^{[a]}=f(uq^{a \hbar})$ for a multiplicative shift ($q$-difference), where 
$a \in {\mathbb C}$. 
Here the unit of the shift $\hbar $ is any non-zero fixed complex number.  
If there is no shift ($a=0$), we often omit $[0]$: $f^{[0]}=f=f(u)$. 
\subsection{QQ-relations}
Let us consider $2^{M+N}$ complex functions 
$\Qb_{I}=\Qb_{I}(u) $ of 
 $u \in {\mathbb C}$ (the spectral parameter), which are   
labeled by subsets $I$ of the full set ${\mathfrak I} $.  
We   use an abbreviation of the form $\Qb_{I,i,j}=
 \Qb_{\{ i_{1},i_{2},\dots, i_{a},i,j\}}
=\Qb_{i_{1},i_{2},\dots, i_{a},i,j}$ 
for $I=\{i_{1},i_{2},\dots, i_{a}\}$. 
We also assume that these functions $\{ \Qb_{I}\}$ 
 satisfy the following functional relations: 
\begin{align}
& (z_{i}-z_{j})\Qb_{I}\Qb_{I,i,j}
=z_{i}\Qb_{I,i}^{[p_{i}]}
\Qb_{I,j}^{[-p_{i}]}-
z_{j}\Qb_{I,i}^{[-p_{i}]}
\Qb_{I,j}^{[p_{i}]}
\qquad 
\text{for} \qquad p_{i}=p_{j},   
\label{QQb}  \\[6pt]
& 
(z_{i}-z_{j})\Qb_{I,i}\Qb_{I,j}=
z_{i}\Qb_{I}^{[-p_{i}]}
\Qb_{I,i,j}^{[p_{i}]}-
z_{j}\Qb_{I}^{[p_{i}]}
\Qb_{I,i,j}^{[-p_{i}]}
\qquad \text{for} \qquad p_{i}=-p_{j}, 
\label{QQf} 
\end{align} 
where $i,j \in {\mathfrak I} \setminus I$ ($i\ne j$); 
$\{z_{i} \}_{i \in {\mathfrak I}}$ are complex parameters. 
In the context of quantum integrable spin chains, the parameters $\{z_{i} \}_{i \in {\mathfrak I}}$ 
correspond to boundary twist parameters, and \eqref{QQb}-\eqref{QQf} are functional relations 
among Baxter Q-functions (or operators), called QQ-relations. 
Then we will call the function $\Qb_{I}$ `Q-function' from now on. 
These relations (their special cases, extensions or variations) appeared 
in various guises in the literature (see for example, 
\cite{BLZ98,PS00,DDT00,MV04,KSZ07}
 and references in \cite{T09}). 
The first equations \eqref{QQb} (Bosonic QQ-relations) are generalization of the quantum Wronskian condition. 
 The second equations \eqref{QQf} (Fermionic QQ-relations) came from 
the particle-hole transformations 
\cite{Wo83} 
 in statistical mechanics, 
 and are related \cite{T98} to odd Weyl reflections \cite{DP85,Se85} of the superalgebra $gl(M|N)$.  
The fermionic QQ-relations are studied in relation to B\"{a}cklund transformations in soliton theory \cite{KSZ07}.  
Here we use the form \cite{T09} defined for the whole set of $2^{M+N}$ Q-functions. 
In \cite{T09,T11}, we normalized the Q-function for the empty set as $\Qb_{\emptyset}=1$, 
but we do not impose this in this paper
\footnote{
However, readers may formally set $\Qb_{\emptyset}=1$ without spoiling the essential structure of the formulas. 
Under this condition, the bulky normalization functions, which appear in what follows,
 become trivial: $\Psi^{(m,n)}_{\mu}=\Psi^{(m,n)}_{a,s}=\Phi_{\mu}=\Phi_{a,s}=1$.}. 
%

We introduce a map $\sigma$, which is related to an automorphism of $gl(M|N)$ (cf. \cite{FSS89}):
\begin{align}
\sigma(z_{i})& =
\begin{cases}
z_{M+1-i}^{-1} 
\quad \text{for} \quad i \in {\mathfrak B}, \\
z_{2M+N+1-i}^{-1} 
\quad \text{for} \quad i \in {\mathfrak F}.
\end{cases}
\label{sigmach}
\end{align}
We also define the action of this map on the index sets of Q-functions as follows.
For any $ \{b_{i}\}_{i=1}^{m} 
\subset {\mathfrak B}, 
\{f_{i}\}_{i=1}^{n} \subset {\mathfrak F}$, 
we define 
\begin{align}
& \sigma ( \{b_{i}\}_{i=1}^{m}  \sqcup \{f_{i}\}_{i=1}^{n} )
={\mathfrak I} \setminus 
(\{ M+1-b_{i}\}_{i=1}^{m} \sqcup \{ 2M+N+1-f_{i} \}_{i=1}^{n} ) .
\label{sigma-set} 
\end{align} 
Then we define the action of the map on the Q-functions by 
\begin{align}
\sigma (\Qb_{\{b_{i}\}_{i=1}^{m}  \sqcup \{f_{i}\}_{i=1}^{n}}) 
= \Qb_{\sigma (\{b_{i}\}_{i=1}^{m}  \sqcup \{f_{i}\}_{i=1}^{n})} 
\label{sigmaQ}.
\end{align} 
Note that  the QQ-relations \eqref{QQb} and \eqref{QQf} 
are invariant 
 under the map $\sigma$. 
 
\subsection{Tableau sum and CBR-type expressions of T-functions}
Let us take a  
 $K$-tuple $I_{K}=(\gamma_{1},\gamma_{2},\dots,\gamma_{K})$ 
whose components are mutually distinct elements of ${\mathfrak I}$, 
where $0 \le K \le M+N$. In particular, $I_{0}$ and $I_{M+N}$ coincide with $\emptyset $ and ${\mathfrak I}$ as sets (rather than tuples),  
respectively. 
Then we define functions
\footnote{For any $K$, we regard the tuple $I_{K}$ as a set $I_{K}=\{\gamma_{1},\gamma_{2},\dots,\gamma_{K}\}$ 
for the Q-function $\Qb_{I_{K}}$ and the summation $\sum_{j \in I_{K}}$.
We also remark that ${\mathcal   X}_{I_{K}}^{[-\frac{M-N}{2}]}$ 
corresponds to eq.(2.7) in \cite{T09}.}
 labeled by $I_{K}$ as 
\begin{align}
{\mathcal   X}_{I_{K}}&:=
z_{\gamma_{K}}
\frac{\Qb_{I_{K-1}}^{[-\sum_{j \in I_{K-1}}p_{j}-2p_{\gamma_{K}}]}
\Qb_{I_{K}}^{[-\sum_{j \in I_{K}}p_{j}+2p_{\gamma_{K}}]}
}{
\Qb_{I_{K-1}}^{[-\sum_{j \in I_{K-1}}p_{j}]}
\Qb_{I_{K}}^{[-\sum_{j \in I_{K}}p_{j}]}
} .
\label{boxes} 
\end{align}
The eigenvalue formula of the transfer matrix of the Perk-Schultz-type model \cite{Perk:1981}  
by  Bethe ansatz has the following form
 \cite{Sc83} (see \eqref{TF-rel03} for more familiar normalization):
\begin{eqnarray}
{\mathcal F}_{(1)}^{I_{M+N}}=
\sum_{a=1}^{M+N}p_{\gamma_{a}}{\mathcal   X}_{I_{a}}^{[M-N]}.
 \label{tab-fund}
\end{eqnarray}
A  partition is a non increasing sequence of positive 
 integers $\mu=(\mu_{1},\mu_{2},\dots) $: 
$\mu_{1} \ge \mu_{2} \ge \dots \ge 0$. 
We often write this in the form $\mu=(r^{m_{r}},(r-1)^{m_{r-1}},\dots,2^{m_{2}},1^{m_{1}})$, 
where $r=\mu_{1}$, and 
 $m_{k}={\rm Card}\{j|\mu_{j}=k \}$. 
We use the same symbol $\mu$ for the Young diagrams. 
The Young diagram $\mu$, corresponding to a partition $\mu$, has 
$\mu_{k}$ boxes in the $k$-th row in the plane.
Each box in the Young diagram is specified by the coordinates  
$(i,j)\in {\mathbb Z}^{2}$, 
where the row index $i$ increases as one goes downwards, and the column 
index $j$ increases as one goes from left to right. 
The top left corner of $\mu$ has the coordinates $(1,1)$. 
The conjugate (transposition) of $\mu$ is defined by $\mu^{\prime}=(\mu_{1}^{\prime},\mu_{2}^{\prime},\dots)$,  
where $\mu_{j}^{\prime}={\rm Card}\{k| \mu_{k} \ge j\}$. 
Let $\lambda =(\lambda_{1},\lambda_{2},\dots)$ and 
$\mu =(\mu_{1},\mu_{2},\dots)$ be two partitions such that
$\mu_{i} \ge \lambda_{i}: i=1,2,\dots$ and 
$\lambda_{\mu_{1}^{\prime}}=\lambda^{\prime}_{\mu_{1}}=0$. 
We express the skew-Young diagram defined by these two partitions as 
 $\lambda \subset \mu$. 
Each box on the skew-Young diagram $\lambda \subset \mu$ is specified by its coordinate on $\mu$. 
We also use the $180$ degrees rotation of the skew-Young diagram 
$\lambda \subset \mu $, and 
denote it as $\widetilde{\lambda \subset \mu}$:
\begin{multline}
\widetilde{\lambda \subset \mu}:=
(\underbrace{\mu_{1}-\mu_{\mu_{1}^{\prime}},\mu_{1}-\mu_{\mu_{1}^{\prime}-1}, 
\mu_{1}-\mu_{\mu_{1}^{\prime}-2},\dots,\mu_{1}-\mu_{2},0}_{\mu_{1}^{\prime}} )
\\
\subset 
(\mu_{1}^{\mu_{1}^{\prime}-\lambda_{1}^{\prime}},\mu_{1}-\lambda_{\lambda_{1}^{\prime}},
\mu_{1}-\lambda_{\lambda_{1}^{\prime}-1},
\mu_{1}-\lambda_{\lambda_{1}^{\prime}-2},
\dots, 
\mu_{1}-\lambda_{2},\mu_{1}-\lambda_{1}).
\label{180deg}
\end{multline}
Note  that $ \widetilde{\mu}=\mu $  if 
$\mu $ is 
 a Young diagram of rectangular shape. 
%
We define the space of admissible tableaux  
$\mathsf{Tab}_{I_{K}}(\lambda\subset \mu)$ 
for the $K$-tuple $I_{K}$  
on a (skew) Young diagram $\lambda\subset \mu$. 
%
We assign an integer $t_{ij}$ in each box $(i,j)$ of the diagram. 
An admissible tableau 
$t\in\mathsf{Tab}_{I_{K}}(\lambda\subset \mu)$ 
 is a set of integers $t=\{t_{jk}\}_{(j,k)\in \lambda\subset \mu}$, 
where all $t_{jk} \in \{1, 2, \dots, K \}$  
satisfy the following conditions  
\begin{itemize} 
\item[(i)] $t_{jk}\le t_{j+1,k},t_{j,k+1}$ \\
\item[(ii)]  $t_{jk} < t_{j,k+1}$ if $\gamma_{t_{jk}}\in {\mathfrak F}$ or
  $\gamma_{t_{j,k+1}} \in {\mathfrak F}$  \\ 
\item[(iii)]  $t_{jk} < t_{j+1,k}$ if $\gamma_{t_{jk}}\in {\mathfrak B}$ or
  $\gamma_{t_{j+1,k}} \in {\mathfrak B}$.
\end{itemize}
We introduce a $\Ts$-function with auxiliary space labeled by 
a skew Young diagram $\lambda\subset \mu$ 
\cite{T97,T98} 
(see \cite{T98-2} for an extension of these T-functions, 
\cite{BR90} for $N=0$ case, \cite{LM20} for representation or combinatorial theoretical background and \cite{KOS95} for $U_{q}(B_{r}^{(1)})$ case):  
\begin{align}
{\mathcal  F}_{\lambda\subset \mu}^{I_{K}}=
\sum_{t \in \mathsf{Tab}_{I_{K}}(\lambda\subset \mu)}
\prod_{(j,k) \in \lambda\subset \mu}
p_{\gamma_{t_{j,k}}}
{\mathcal  X}_{I_{t_{j,k}}}^{[\mu_{1}-\mu_{1}^{\prime}+2j-2k+m-n]},
\label{DVF-tab1} 
\end{align}
where the summations are taken over all admissible tableaux, and 
the products are taken over all boxes of the 
Young diagram $\lambda\subset \mu$;   
$m:={\rm card} (I_{K} \cap {\mathfrak B})$, 
$n:={\rm card} (I_{K} \cap {\mathfrak F})$. 
We also set ${\mathcal  F}_{\emptyset}^{I_{K}}=1$, 
and 
${\mathcal  F}_{\mu}^{\emptyset}=0$ and 
for the non-empty Young diagram $\mu $.
Note that the admissible tableaux $\mathsf{  Tab}_{I_{K}}(\lambda \subset \mu)$ 
 becomes an empty set if the Young diagram $ \lambda \subset \mu$ 
contains a rectangular sub-diagram of a height of $m+1$ and a width of $n+1$,
 and consequently
(\ref{DVF-tab1})
 vanishes for such Young diagram. Thus the T-functions for $\lambda = \emptyset $ are defined on $[m,n]$-hook (L-hook). 

The tableau sum formula \eqref{DVF-tab1} has a determinant expression.
\begin{align}
{\mathcal F}_{\lambda \subset \mu}^{I_{K}}&= 
\begin{vmatrix}
    \left(
    {\mathcal F}^{I_{K} \, [\mu_{1}-\mu_{1}^{\prime}+\mu_{i}^{\prime}+\lambda_{j}^{\prime}-i-j+1]}_{(1^{\mu_{i}^{\prime}-\lambda_{j}^{\prime}-i+j})} 
    \right)_{1 \le i \le \mu_{1} \atop
    1 \le j \le \mu_{1}}
 \end{vmatrix}   
 \label{superJT1}
\end{align}
where  ${\mathcal F}_{(1^{0})}^{I_{K}}={\mathcal F}_{(0)}^{I_{K}}=1$ and 
${\mathcal F}_{(1^{a})}^{I_{K}}={\mathcal F}_{(a)}^{I_{K}} = 0$ for $a <0$. 
This determinant expression for $K=M+N$ corresponds to 
the supersymmetric Cherednik-Bazhanov-Reshetikhin formula (supersymmetric CBR formula or quantum supersymmetric Jacobi-Trudi formula) \cite{T97,T98} 
(see also \cite{KV07,LM20,KOS95}), which is 
a supersymmetric extension of the CBR formula (quantum Jacobi-Trudi formula) \cite{Ch89,BR90}. 
We introduce the following transformation for any (non-skew) Young diagram $\mu$.  
\begin{align}
 {\mathsf F}_{\mu}^{I_{K}}
 =
\Qb_{\emptyset}^{[m-n+\mu_{1}-\mu_{1}^{\prime}]} 
\Qb_{I_{K}}^{[-\mu_{1}+\mu_{1}^{\prime}]}
{\mathcal  F}_{\widetilde{\mu}}^{I_{K}}.
\label{TF-rel03}
\end{align}
Here  $\widetilde{\mu}$ is 
 the 180 degrees rotated Young diagram of $\mu$ (see, \eqref{180deg} for $\lambda=\emptyset$). 
 It is known that ${\mathcal F}_{\mu}^{I_{K}}$ and ${\mathsf F}_{\mu}^{I_{K}}$ are invariant under any permutation of the elements of 
 the tuple $I_{K}$ under the QQ-relations \eqref{QQb} and \eqref{QQf}. 
 Then we define ${\mathcal F}_{\mu}^{B,F}:={\mathcal F}_{\mu}^{I_{K}}$ and 
 ${\mathsf F}_{\mu}^{B,F}:={\mathsf F}_{\mu}^{I_{K}}$, where 
 $B=I_{K} \cap \Bm$ and $F=I_{K} \cap \Fm$ as sets. 
 \subsection{Wronskian-type expressions of T-functions}
Let us introduce a determinant 
\footnote{
This is related to the sparse determinant in [eq.(3.24) in \cite{T11}] as
$
\Delta^{B,R
,[\xi ]}_  
{F,S  
} 
=
\Delta^{B,\emptyset,R
,[0; \xi ]}_  
{F,S, \emptyset, \emptyset  
} 
$,
where  we set $B_{1}=B, B_{2}=T_{1}=T_{2}=\emptyset$, $\eta =0$.
Moreover, this is related to the determinant in [eq.(3.4) in \cite{T09}] 
as 
$\Delta^{B,R
,[\xi ]}_  
{F,S  
} 
=\Delta^{B,R}_  
{F,S  
}(x q^{\xi}) $ in case q-difference is adopted. 
}
over a block matrix 
labeled by sets $B,R,F,S$ ($B \subset \Bm$, $|B|=m$; $F \subset \Fm$, $|F|=n$; $R,S \subset \mathbb{Z}$): 
\begin{align}  
\Delta^{B,R,[\xi ]}_  
{F,S  
} 
&=   
\begin{vmatrix}  
\left( \frac{\Qb_{b,f}^{[\xi ]}}{z_{b}- z_{f} } 
\right)_{  
\genfrac{}{}{0pt}{}{b\in B, }{f \in F} } 
&   
 \left(
z_{b }^{j-1} \Qb_{b}^{[\xi+2j-1]}
\right)_{  
\genfrac{}{}{0pt}{}{b\in B, }{j \in S} }    
  \\[6pt]  
\left(
(-z_{f})^{i-1}\Qb_{f}^{[\xi-2i+1 ]}
\right)_{  
\genfrac{}{}{0pt}{}{i \in R, }{f \in F} }  
& (0)_{|R|\times |S|}    
\end{vmatrix}  
,  \label{sparcedetbb}
\end{align}  
where  $\xi \in \mathbb{C}$,  and $(0)_{|R|\times |S|} $ is $|R|$ by $ |S|$ zero matrix.  
The number of the elements of the sets must satisfy 
$|B|+|R|=|F| + |S| $.
For any Young diagram $\mu$, we introduce a 
number, called $(m,n)$-index \cite{MV03}:
\begin{align}
\xi_{m,n}(\mu):={\rm min}\{j \in {\mathbb Z}_{\ge 1}|\mu_{j}+m-j \le n-1\}.
\label{mn-index}
\end{align}
In particular, we have $1 \le \xi_{m,n}(\mu) \le m+1$, $\xi_{m,0}(\mu)=m+1$ and $\xi_{0,n}(\mu)=1$
 for $\mu_{m+1} \le n$, and 
$\xi_{m,n}(\mu)=m+1$ for $\mu_{m+1} \le n \le \mu_{m}$; 
$\xi_{m,n}(\emptyset)=\max\{m-n+1,1\}$. 
We often abbreviate $\xi_{m,n}(\mu)$ as $\xi_{m,n}$. 
The denominator formula of the supercharacter of $gl(m|n)$ can be written as
 \cite{MV03}: 
\begin{align}
 \Ds( B|F)
&= 
\frac{\prod_{ b,b^{\prime } \in B, \atop b < b^{\prime } }(z_{b} - z_{b^{\prime}})
\prod_{f,f^{\prime } \in F, \atop f < f^{\prime } }(z_{f^{\prime} } - z_{f})
}
{\prod_{ (b,f) \in B \times F } (z_{b} - z_{f})}.
\end{align} 

For any Young diagram $\mu$, 
we introduce the following function
\footnote{$ \Ts_{\mu}^{B,F [-\frac{m-n}{2}]}$ 
corresponds to eq.(3.15) in \cite{T09}.}
\begin{multline}
 \Ts_{\mu}^{B,F}
:= (-1)^{(m+n+1)(\xi_{m,n}(\mu ) +1) +\frac{(m-n)(m+n-1)}{2}} 
 \\ 
 \times 
 \Psi^{(m,n)}_{\mu}
\Delta^{B, (r_{1},r_{2},\dots,r_{n-m+\xi_{m,n}-1}), [-m+n +\mu_{1}^{\prime}-\mu_{1}]}
_{F,(s_{1},s_{2},\dots,s_{\xi_{m,n}-1} )} 
 \Ds( B|F)^{-1},
\label{unnor-t1}
\end{multline}
where $s_{l}=\mu_{\xi_{m,n} -l}+m-n-\xi_{m,n}(\mu)+l+1$,
$r_{k}=\mu_{n-m+\xi_{m,n}-k}^{\prime}+k-\xi_{m,n}(\mu)+1$ and 
\begin{align}
 \Psi^{(m,n)}_{\mu}=
 \frac{\Qb_{\emptyset}^{[m-n+\mu_{1}-\mu_{1}^{\prime}]} \Qb^{[-m+n-\mu_{1}+\mu_{1}^{\prime}]}_{\emptyset}
 \left( \Qb_{\emptyset}^{[-m+n-\mu_{1}+\mu_{1}^{\prime}]}\right)^{-m-1+\xi_{m,n}}}
{
\prod^{n-m+\xi_{m,n}-1}_{i=1} \Qb_{\emptyset}^{[-m+n-\mu_{1}+\mu_{1}^{\prime}-2r_{i}+2]}
\prod^{\xi_{m,n}-1}_{j=1} \Qb_{\emptyset}^{[-m+n-\mu_{1}+\mu_{1}^{\prime}+2s_{j}-2]} 
 } .
 \label{Del-nor}
\end{align}
We remark that \eqref{unnor-t1} reduces to a determinant formula \cite{MV03} of characters of tensor representations of $gl(m|n)$ 
 in the limit (or formal replacement) $\Qb_{\emptyset} \to 1$, $\Qb_{b} \to 1$, $\Qb_{f} \to 1$, $\Qb_{b,f} \to 1$. 
Specializing \eqref{unnor-t1} for the empty diagram, we obtain 
a Wronskian-like determinant solution [Theorem 3.2 in \cite{T09}] of the QQ-relations \eqref{QQb} and \eqref{QQf}: 
\begin{align}
 \Ts_{\emptyset}^{B,F}
& =
(-1)^{\frac{(m-n)(m+n-1)}{2}}  \Psi^{(m,n)}_{\emptyset}
\Delta^{B, \langle 1, n-m \rangle , [-m+n ]}
_{F, \langle 1, m-n \rangle } 
 \Ds( B|F)^{-1}
 \nonumber 
 \\
 &=\Qb_{B,F} \Qb_{\emptyset}^{[m-n]},
 \label{emptyT}
\end{align}
where we introduce notation
\begin{align}
\langle a,b \rangle =
\begin{cases}
\{a,a+1,a+2, \dots, b \} & \text{for} \quad b -a \in {\mathbb Z}_{\ge 0}, \\[3pt]
\emptyset & \text{for}  \quad  b -a  \notin {\mathbb Z}_{\ge 0} .
\end{cases}
\end{align}
For any rectangular Young diagram, we set
\begin{align}
\Ts_{a,s}^{B,F}&=
\begin{cases}
\Ts^{B,F}_{(s^{a})} & \text{for} \quad a, s \in \mathbb{Z}_{\ge 1}
\\
\Qb_{\emptyset}^{[m-n-a]} (\Qb_{\emptyset}^{[m-n+a]})^{-1} \Ts^{B,F[a]}_{\emptyset} & \text{for} \quad a \in \mathbb{Z}_{\ge 0}, \quad s  =0
\\
\Qb_{\emptyset}^{[m-n+s]} (\Qb_{\emptyset}^{[m-n-s]})^{-1} 
\Ts^{B,F[-s]}_{\emptyset} & \text{for} \quad  a=0 , \quad s \in \mathbb{Z}
\\
0  & \text{otherwise}  .
\end{cases}
 \label{Trec0}
\end{align}
More explicitly, we have:
\\
\noindent 
for $a \le m-n$, we have $\xi_{m,n}((s^a))=m-n+1$ and 
\begin{align}
 \Ts_{a,s}^{{B,F}}
= 
(-1)^{\frac{(m-n)(m+n-1)}{2}} \Psi^{(m,n)}_{a,s}\Delta^{B, \emptyset , [-m+n+a-s]}
_{F,\langle 1,m-n-a \rangle  \sqcup \langle m-n-a+s+1,m-n+s \rangle}
 \Ds( B|F)^{-1},
  \label{Trec1}
\end{align}
for $a-s \le m-n \le a $, we have $\xi_{m,n}((s^a))=a+1$ and 
\begin{align}
\Ts_{a,s}^{B,F}
= (-1)^{(m+n+1)a+\frac{(m-n)(m+n-1)}{2}} \Psi^{(m,n)}_{a,s}
\Delta^{B, \langle 1, n-m+a \rangle , [-m+n+a-s]}
_{F,\langle m-n-a+s+1 , m-n+s \rangle }
 \Ds( B|F)^{-1},
  \label{Trec2}
\end{align}
for $-s \le m-n \le a-s $, we have $\xi_{m,n}((s^a))=m-n+s+1$ and 
\begin{align}
\Ts_{a,s}^{B,F}
= (-1)^{(m+n+1)s+\frac{(m-n)(m+n-1)}{2}} \Psi^{(m,n)}_{a,s}
\Delta^{B,\langle n-m-s+a+1,n-m+a \rangle , [-m+n+a-s]}
_{F,\langle 1, m-n+s \rangle}
 \Ds( B|F)^{-1},
 \label{Trec3}
\end{align}
for $m-n \le -s$, we have $\xi_{m,n}((s^a))=1$ and 
\begin{align}
\Ts_{a,s}^{B,F}
= (-1)^{\frac{(m-n)(m+n-1)}{2}} \Psi^{(m,n)}_{a,s}
\Delta^{B,\langle 1,n-m-s \rangle \sqcup \langle n-m-s+a+1,n-m+a \rangle , [-m+n+a-s]}
_{F, \emptyset}
 \Ds( B|F)^{-1},
  \label{Trec4}
\end{align}
where we set
\begin{align}
\Psi^{(m,n)}_{a,s}&=
\begin{cases}
\Psi^{(m,n)}_{(s^{a})} & \text{for} \quad a, s \in \mathbb{Z}_{\ge 1}
\\
\Qb_{\emptyset}^{[m-n-a]} (\Qb_{\emptyset}^{[m-n+a]})^{-1} \Psi^{(m,n) [a]}_{\emptyset} & \text{for} \quad a \in \mathbb{Z}_{\ge 0}, \quad s  =0
\\
\Qb_{\emptyset}^{[m-n+s]} (\Qb_{\emptyset}^{[m-n-s]})^{-1} 
\Psi^{(m,n) [-s]}_{\emptyset} & \text{for} \quad  a=0 , \quad s \in \mathbb{Z}
\\
1  & \text{otherwise}  .
\end{cases}
 \label{Tas-nor}
\end{align}
Applying the Laplace expansion formula to \eqref{Trec0}-\eqref{Trec4}, we obtain useful expressions \cite{T09}
\begin{multline}    
\Ts^{B,F}_{a,s} =   
\sum_{I\sqcup J= F,  \atop
|I|=n-s}  
\frac{\prod_{j \in J}(-z_{j})^{a-s-m+n} \prod_{(b,j) \in B \times J } (z_{b}-z_{j}) }
{\prod_{(i,j) \in I \times J  } (z_{i}-z_{j}) }
\Qb^{[a]}_{B,I}  
\Qb^{[-a+m-n]}_{J}  
\\
 \qquad \qquad  \text{for} \quad a\ge s+m-n,  
\label{soQQb-1}  
\end{multline}
\begin{multline}
\Ts^{B,F}_{a,s} =   
\sum_{I\sqcup J=B, \atop |I|=a} 
\frac{ \prod_{i \in I} z_{i}^{s-a+m-n}
\prod_{(i,f) \in I \times F } (z_{i}-z_{f})}
{\prod_{(i,j) \in I \times J  } (z_{i}-z_{j}) }  
\Qb^{[s+m-n]}_{I}  
\Qb^{[-s]}_{J, F}  
\quad 
\text{for} \quad  a \le s + m-n, \label{soQQb-2}  
\end{multline} 
where the summation is taken over any possible decomposition of the original set 
into two disjoint sets $I $ and $J$ with fixed sizes. 
The T-functions $\Ts^{B,F}_{a,s}$ solve \cite{T09} the T-system for $U_{q}(gl(M|N)^{(1)})$ (or $U_{q}(sl(M|N)^{(1)})$,  
or its Yangian counterpart) \cite{T97,T98}. 
We also have [eq.(3.67) in  \cite{T09}].
\begin{align}
\Ts^{B,F}_{\mu}={\mathsf F}^{B,F}_{\mu} 
 \label{T=F}
\end{align}
This is proven for general non-rectangular Young diagrams for $|F| =0$ case and rectangular Young diagrams for $|B||F| \ne 0$ case, 
and is a conjecture for general non-rectangular Young diagrams for $|B||F| \ne 0$ case (see page 426 in \cite{T09}).
\section{T-and Q-functions for $U_{q}(B^{(1)}_{r})$}
Now we would like to explain details of our proposal [section 3.7 in \cite{T11}]. 
In this section we consider reductions of T- and Q-functions introduced in the previous section. 
We focus on $U_{q}(B^{(1)}_{r})$ case, and give Wronskian solutions of the T-system. 
\subsection{Reductions of Q-functions by automorphisms}
We find that reductions on the  QQ-relations by the map $\sigma$ 
 (and some dualities among different superalgebras  \cite{Z97}) 
 produce QQ-relations (and from zeros of Q-functions,  
Bethe equations) and solutions of the T-systems associated with algebras different from the original ones. 
The reductions here are basically accomplished by identifying the image of 
the Q-functions and the parameters $\{z_{a}\}$ by the map $\sigma$ 
with the original ones (up to overall factors and manipulations on the spectral parameter in some cases). 
Let us consider `$gl(M|N)^{(2)}$ type reduction'
\footnote{In the previous paper \cite{T11}, we used the term `$sl(M|N)^{(2)}$ type reduction'. 
Here we change this since we do not impose the special linear algebra condition explicitly.} by $\sigma$: 
\begin{align}
\sigma (\Qb_{I})=\Qb_{I} \quad \text{for} \quad I \subset {\mathfrak I}, \quad 
\sigma (z_{a})=z_{a} \quad \text{for} \quad a \in {\mathfrak I} . 
\label{reduction-sigma}
\end{align} 
If $M$ or $N$ are odd, fixed points by $\sigma$ appear. 
For example for $N=2r+1$ case, we have 
$\sigma(z_{M+r+1})=z_{M+r+1}^{-1}=z_{M+r+1}$. Then we obtain $z_{M+r+1}=\pm 1$. 
The minus sign $z_{M+r+1}=-1$ effectively changes the sign of $p_{M+r+1}$ from the grading of the superalgebra 
(see \eqref{boxes} and \eqref{tab-fund}), 
which induces a duality (some kind of Boson-Fermion correspondence) 
between a superalgebra ($z_{M+1}=1$) and an ordinary (non-graded) algebra ($z_{M+1}=-1$) for the case $N=1$. 
In particular, $gl(2r|1)^{(2)}$ type reduction
\footnote{This was announced first at the poster session in the conference 
`Infinite Analysis 09, New Trends in Quantum Integrable Systems', 
Department of Mathematics, Faculty of Science, 
Kyoto University, 29 July 2009.} 
 for $z_{2r+1}=-1$, which is the main subject of this paper,  
 produces QQ-relations  and Wronskian solutions 
of the T-system for $U_{q}({\mathfrak g}^{(1)})$ or $Y({\mathfrak g})$, where ${\mathfrak g}=so(2r+1)$. 
Moreover, 
$gl(0|2r+1)^{(2)}$ type reduction for $z_{r+1}=-1$ corresponds to ${\mathfrak g}=osp(1|2r)$ (cf.\ \cite{T99}). 
$gl(2r+1|0)^{(2)}$ type reduction
\footnote{In this case, we have to add a shift of the spectral parameter to \eqref{reduction-sigma} as 
$\sigma (Q_{I})=Q_{I}^{[\frac{\pi i}{2\hbar}]}  $ ($\hbar \in {\mathbb C} \setminus \{0 \}$ 
cf. \cite{KS94-2,T02}).  
}
 for $z_{r+1}=1$ corresponds to $U_{q}(A_{2r}^{(2)})$, and 
$gl(2r|0)^{(2)}$ type reduction  corresponds to $U_{q}(A_{2r-1}^{(2)})$ (cf. \cite{T02,KS94-2}).  
These reductions produce additional functional relations among T-functions, which do not exist before the reductions 
(see for example, \eqref{T+T}). 
\subsection{QQ-relations}
In this subsection, we will show that the QQ-relations for quantum integrable systems associated with $U_{q}(B_{r}^{(1)})=U_{q}(so(2r+1)^{(1)})$ 
follow from the QQ-relations \eqref{QQb} and \eqref{QQf} by the  reduction \eqref{reduction-sigma}. 
 
Now we consider the case $M=2r$, $N=1$, $\Bm=\{1,2,\dots , 2r\}$, $\Fm=\{2r+1\}$. 
For $a \in \Bm $ and  $B \subset \Bm $, set $a^{*}=2r-a+1$,  $B^{*}=\{a^{*}| a \in B \}$.
In this case
\footnote{The case $z_{2r+1}=1$, $U_{q}(gl(2r+1|1)^{(2)})$ is parallel to this.}
, \eqref{reduction-sigma} reduces to 
\begin{align}
&z_{a^{*}}=z_{a}^{-1} \quad \text{for} \quad a \in \Bm,
\qquad 
z_{2r+1}=-1, 
\label{z-z}
\\
& \Qb_{B}=\Qb_{\Bm \setminus B^{*}, \Fm}
\quad \text{for} \quad B \subset \Bm . 
\label{Q-Q}
\end{align}
In particular, we have
\begin{align}
\prod_{j \in \Bm}z_{j}=1 ,
\label{proz=1}
\end{align}
and 
\begin{align}
&\Qb_{\Bm}=\Qb_{\Fm}=\Qb_{2r+1} , \label{Q=Q}
\\
&\Qb_{\Bm,\Fm}=\Qb_{\emptyset} . \label{Q=Q2}
\end{align}
Let us take a subset of \eqref{QQb}.
\begin{align}
& (z_{i}-z_{j})\Qb_{B}\Qb_{B,i,j}
=z_{i}\Qb_{B,i}^{[1]}
\Qb_{B,j}^{[-1]}-
z_{j}\Qb_{B,i}^{[-1]}
\Qb_{B,j}^{[1]} ,
\label{QQbten}
\end{align} 
where $B \subset \Bm$, $i,j \in \Bm \setminus B$, $i \ne j$.
 We find that Q-functions satisfy 
\begin{align}
& (z_{b}^{\frac{1}{2}}-z_{b^{*}}^{\frac{1}{2}})\Qb^{[\frac{1}{2}]}_{I}\Qb^{[-\frac{1}{2}]}_{I}
=z_{b}^{\frac{1}{2}} \Qb_{I,b}^{[\frac{1}{2}]}
\Qb_{I,b^{*}}^{[-\frac{1}{2}]}-
z_{b^{*}}^{\frac{1}{2}} \Qb_{I,b}^{[-\frac{1}{2}]}
\Qb_{I,b^{*}}^{[\frac{1}{2}]}
,   
\label{QQsp}  
\end{align} 
where the set $I$ obeys the following conditions: $I \subset \Bm$, $\mathrm{Card}(I)=r-1$,  $i^{*} \notin I$ for any $i \in I$, 
and 
  $b, b^{*} \in \Bm \setminus I$ (thus $\Bm=I \sqcup I^{*} \sqcup \{b,b^{*}\}$). 
  QQ-relations corresponding to \eqref{QQbten} for $B=\{1,2,\dots, i-1\}$, $j=i+1 \le r$,  and \eqref{QQsp} for 
  $I=\{1,2,\dots, r-1\}$, $b=r$ and $b^{*}=r+1$ appeared in [eq.\ (6.11) and eq.\ (6.14) in \cite{DDMST06}].
  \footnote{We may have more Q-functions than the case we consider the type B-system from the very beginning. The extra Q-functions 
  we have play an auxiliary role in our construction of  Wronskian expressions of T-functions. 
  How the extra Q-functions and QQ-relations (if any) affect the class of solutions of the type B T-system remains to be clarified. 
  In relation to this, we made the following experiments by Mathematica version 7. 
 (i) Let us consider the determinant 
$\Qb_{B}=\Ts_{\emptyset}^{B,\emptyset}/\Qb_{\emptyset}^{[m]}$ \eqref{emptyT}, which automatically solves \eqref{QQbten}. 
Substituting this into \eqref{QQsp}, we obtain functional relations among the Q-functions $\{ \Qb_{b}\}_{b \in \Bm}$ and $\Qb_{\emptyset}$. 
Then we solve these order by order by series expansions $\Qb_{b}=1+\sum_{k=1}^{\infty}a_{b}^{(k)} u^{k}$, 
$\Qb_{\emptyset }=1+\sum_{k=1}^{\infty}a_{\emptyset}^{(k)} u^{k}$. 
Here we regard the Q-functions as functions of the spectral parameter $u$ and adopt multiplicative shift
 $\Qb^{[a]}_{I}=\Qb^{[a]}_{I}(u)=\Qb_{I}(q^a u)$ for any $I \subset \Im $. The Q-function $\Qb_{\emptyset}$ is regarded 
 as a known function. 
(ii) Similarly, substituting the determinant \eqref{emptyT} into the relation \eqref{Q-Q}, we obtain functional relations among  
the Q-functions $\{ \Qb_{b}\}_{b \in \Bm}$, $\Qb_{2r+1}$ , $\{ \Qb_{b,2r+1}\}_{b \in \Bm}$ and $\Qb_{\emptyset}$. 
We solve these order by order by series expansions $\Qb_{b}=1+\sum_{k=1}^{\infty}a_{b}^{(k)} u^{k}$, 
$\Qb_{2r+1}=1+\sum_{k=1}^{\infty}a_{2r+1}^{(k)} u^{k}$, 
$\Qb_{b,2r+1}=1+\sum_{k=1}^{\infty}a_{b,2r+1}^{(k)} u^{k}$, 
$\Qb_{\emptyset }=1+\sum_{k=1}^{\infty}a_{\emptyset}^{(k)} u^{k}$. 
Here the coefficients $\{a_{b,2r+1}^{(k)}\}_{b \in \Bm}$ are expressed in terms of  $\{a_{b}^{(j)}\}_{b \in \Bm}$,
 $a_{2r+1}^{(j)}$, $a_{\emptyset}^{(j)}$, ($1 \le j \le k$) through \eqref{QQf}. 
We have confirmed that both (i) and (ii) give the same coefficients $\{a_{b}^{(k)}\}_{b \in \Bm}$ for $k=1,2$, $r=2,3$ cases. 
 }
Let us rewrite \eqref{QQsp} as 
\begin{align}
& (z_{b}-1)\Qb^{[1]}_{I}\Qb_{I}
=z_{b} \Qb_{I,b}^{[1]}
\Qb_{I,b^{*}}- \Qb_{I,b}\Qb_{I,b^{*}}^{[1]}
,   
\label{QQsp2}
\end{align}
where \eqref{z-z} is used. 
We will use the following QQ-relations:
\begin{align}
& (z_{b}-z_{b}^{-1})\Qb_{I}\Qb_{I,b,b^{*}}
=z_{b} \Qb_{I,b}^{[1]} \Qb_{I,b^{*}}^{[-1]}-z_{b}^{-1} \Qb_{I,b}^{[-1]}\Qb_{I,b^{*}}^{[1]},
 \label{QQb2}
\\
& (z_{b}^{-1}+1)\Qb_{I,b^{*}}\Qb_{I,b,b^{*}}
=z_{b}^{-1} \Qb_{I}^{[-1]} \Qb_{I,b^{*}}^{[1]}+ \Qb_{I}^{[1]}\Qb_{I,b^{*}}^{[-1]}
\label{QQf2}
\\
& (z_{b}+1)\Qb_{I,b}\Qb_{I,b,b^{*}}
=z_{b} \Qb_{I}^{[-1]} \Qb_{I,b}^{[1]}+ \Qb_{I}^{[1]}\Qb_{I,b}^{[-1]} ,
\label{QQf3}
\end{align}
which follow from   \eqref{QQb},  \eqref{QQf}, \eqref{z-z} and  \eqref{Q-Q} 
(in particular, $\Qb_{I,2r+1}=\Qb_{I,b,b^{*}}$, 
$\Qb_{I,b^{*},2r+1}=\Qb_{I,b^{*}}$,  
$\Qb_{I,b,2r+1}=\Qb_{I,b}$).  
Now we prove \eqref{QQsp2} step by step as follows. 
\begin{multline}
[\text{left hand side of \eqref{QQsp2}}] \times 
(z_{b}-z_{b}^{-1})\Qb_{I,b,b^{*}}
=(z_{b}-1)\Qb^{[1]}_{I}
\underbrace{(z_{b}-z_{b}^{-1}) \Qb_{I} \Qb_{I,b,b^{*}}}_{\text{apply \eqref{QQb2}}}
  \\
=(z_{b}-1)\Qb^{[1]}_{I} 
(z_{b} \Qb_{I,b}^{[1]} \Qb_{I,b^{*}}^{[-1]}-z_{b}^{-1} \Qb_{I,b}^{[-1]}\Qb_{I,b^{*}}^{[1]}), 
\label{proofQQ}
\end{multline}
\begin{multline}
[\text{right hand side of \eqref{QQsp2}}] \times 
(z_{b}-z_{b}^{-1})\Qb_{I,b,b^{*}}=
\\
=(z_{b}-z_{b}^{-1})
 (z_{b} \Qb_{I,b}^{[1]} 
\underbrace{ \Qb_{I,b^{*}}\Qb_{I,b,b^{*}}}_{\text{apply \eqref{QQf2}}}
- \underbrace{\Qb_{I,b} \Qb_{I,b,b^{*}}}_{\text{apply \eqref{QQf3}}} \Qb_{I,b^{*}}^{[1]})
  \\
=(z_{b}-z_{b}^{-1})
 \bigl(z_{b} \Qb_{I,b}^{[1]} 
(z_{b}^{-1}+1)^{-1} (z_{b}^{-1} \Qb_{I}^{[-1]} \Qb_{I,b^{*}}^{[1]}+ \Qb_{I}^{[1]}\Qb_{I,b^{*}}^{[-1]})
\\
- 
(z_{b}+1)^{-1}
(z_{b} \Qb_{I}^{[-1]} \Qb_{I,b}^{[1]}+ \Qb_{I}^{[1]}\Qb_{I,b}^{[-1]})
\Qb_{I,b^{*}}^{[1]}\bigr)
\\
=[\text{right hand side of \eqref{proofQQ}}] .
\end{multline}
Hence \eqref{QQsp2} holds since 
$(z_{b}-z_{b}^{-1})\Qb_{I,b,b^{*}}$ is not identically zero.  
One can also show that 
\eqref{QQf2} and \eqref{QQf3} follow from \eqref{QQsp2} and \eqref{QQb2}. 

In closing this subsection, 
we remark that the T-function \eqref{DVF-tab1} for the tuple
 $I_{2r+1}=(1,2,\dots , r) \sqcup (2r+1) \sqcup (r+1,r+2,\dots, 2r)
 =(1,2,\dots , r, 2r+1, r+1,r+2,\dots, 2r)$ 
corresponds to the T-function  for $U_{q}(B^{(1)}_{r})$ [eq.(3.2) in \cite{KOS95}]
\footnote{$I_{2r+1}$ corresponds to $(1,2,\dots, r,0,\overline{r},\dots, \overline{2},\overline{1})$ in the notation in \cite{KOS95}.}
 under the reduction \eqref{z-z}-\eqref{Q-Q}.  
\subsection{Wronskian solutions of the T-system}
Now we would like to introduce T-functions. The T-functions for $U_{q}(B_{r}^{(1)})$ are 
defined on $[2r,1]$-hook. This hook corrupts
\footnote{in the sense of B\"{a}cklund transformations \cite{KSZ07}}
 to $[2r,0]$-hook for the T-functions for spinorial representations. 
The main target domain for the T-system is $[r,0]$-hook in these hooks. The rest of the domain plays an auxiliary role. 

Let us introduce the following functions:
\begin{align}
\Tb_{a,s}&= 
\Phi_{a,s}
\Ts^{\Bm,\Fm}_{a,s} 
\nonumber \\
&= (-1)^{r } \Phi_{a,s} \Psi^{(2r,1)}_{a,s}
\frac{
\Delta^{\Bm,\emptyset ,  
[a-s-2r+1]}_  
{\Fm,  
\langle 1, -a+2r-1 \rangle 
\sqcup
\langle s-a+2r, s+2r-1
\rangle}  
}{\Ds (\Bm|\Fm)}
\quad \text{for} \quad 0 \le a \le r-1, \quad s \ge 0
 , 
 \label{solB1}
\\
\Tb_{r,2s}&=\Phi_{r,2s} \Ts^{\Bm,\Fm}_{r,s} 
%
= (-1)^{r }
\Phi_{r,2s} \Psi^{(2r,1)}_{r,s}
\frac{
\Delta^{\Bm,\emptyset ,  
[-r-s+1]}_  
{\Fm,  
\langle 1, r-1 \rangle 
\sqcup
\langle s+r, s+2r-1
\rangle}  
}{\Ds (\Bm |\Fm )}
\quad  \text{for} \quad s \ge 0  ,
\label{solB2}
\\
\Tb_{r,2s+1}&= \Phi_{r,2s+1}
\prod_{j=1}^{r}(z_{j}^{\frac{1}{2}}+z_{j}^{-\frac{1}{2}})\Ts^{\Bm,\emptyset \, [-\frac{1}{2}]}_{r,s} 
\nonumber \\
&= (-1)^{r }
\Phi_{r,2s+1} \Psi^{(2r,0)[-\frac{1}{2}]}_{r,s}
 \prod_{j=1}^{r}(z_{j}^{\frac{1}{2}}+z_{j}^{-\frac{1}{2}})
\frac{
\Delta^{\Bm,\emptyset ,  
[-r-s-\frac{1}{2}]}_  
{\emptyset,  
\langle 1, r \rangle 
\sqcup
\langle s+r+1, s+2r
\rangle}  
}{\Ds (\Bm|\emptyset)}
\quad \text{for} \quad s \ge 0 ,  
\label{solB3}
\end{align}
where the normalization functions are defined by
\begin{align}
\begin{split}
\Phi_{a,s}&=\frac{\prod_{j=1}^{s}\Qb_{\emptyset}^{[2r-s-a+2j-1]} \Qb_{\Bm,\Fm}^{[a+s-2j]}}
{\Qb_{\emptyset}^{[2r+s-a-1]}  \Qb_{\Bm,\Fm}^{[a-s]} } 
\quad \text{for} \quad 0 \le a \le r-1, \quad s \ge 0,
\\
\Phi_{r,2s}&=\frac{\prod_{j=1}^{s}\Qb_{\emptyset}^{[r-s+2j-1]} \Qb_{\Bm,\Fm}^{[r+s-2j]}}
{\Qb_{\emptyset}^{[r+s-1]}  \Qb_{\Bm,\Fm}^{[r-s]} } \quad \text{for} \quad s \ge 0,
\\
\Phi_{r,2s+1}&=\frac{\prod_{j=1}^{s}\Qb_{\emptyset}^{[r+s-2j+\frac{3}{2}]} \Qb_{\Bm,\Fm}^{[r+s-2j+\frac{1}{2}]}}
{\Qb_{\emptyset}^{[r+s-\frac{1}{2}]}}  \quad \text{for}  \quad s \ge 0.
\end{split}
\end{align}
The $(m,n)$-indexes \eqref{mn-index} for \eqref{solB1}, \eqref{solB2} and \eqref{solB3} are $2r$, $2r$ and $2r+1$, respectively.  
Thus the functions \eqref{Tas-nor} reduce to
\begin{align}
\begin{split}
&\Psi_{a,s}^{(2r,1)}=\frac{\Qb_{\emptyset}^{[2r-1-a+s]}}
{\prod_{j=1}^{2r-1-a}\Qb_{\emptyset}^{[-2r-1+a-s+2j]} \prod_{k=2r-a}^{2r-1}\Qb_{\emptyset}^{[-2r-1+a+s+2k]}} 
\quad \text{for} \quad 0 \le a \le r-1,
\\
&\Psi_{r,s}^{(2r,0)[-\frac{1}{2}]}=\frac{\Qb_{\emptyset}^{[r+s-\frac{1}{2}]} \Qb_{\emptyset}^{[-r-s-\frac{1}{2}]}}
{\prod_{j=1}^{r}\Qb_{\emptyset}^{[-r-s+2j-\frac{5}{2}]} \prod_{j=r+1}^{2r}\Qb_{\emptyset}^{[-r+s+2j-\frac{5}{2}]} } ,
\end{split}
\end{align}
where $s \in {\mathbb Z}_{\ge 0}$. 
The Q-function $\Qb_{\Bm,\Fm}$ in the above expressions should be identified with $\Qb_{\emptyset}$ (see \eqref{Q=Q2}), 
but we keep it so that one can keep track of $U_{q}(gl(2r|1)^{(1)})$ origin of the formulas. 
We remark that the T-function \eqref{solB3} for $s=0$ is proportional to the Q-functions. 
\begin{align}
\Tb_{r,1}=\prod_{j=1}^{r}(z_{j}^{\frac{1}{2}}+z_{j}^{-\frac{1}{2}}) \Qb_{\Bm}^{[r-\frac{1}{2}]}
=\prod_{j=1}^{r}(z_{j}^{\frac{1}{2}}+z_{j}^{-\frac{1}{2}}) \Qb_{2r+1}^{[r-\frac{1}{2}]}
\label{Tr1}
\end{align}
We find \cite{T11} that \eqref{solB1}-\eqref{solB3} solves the T-system of type B
 \cite{KNS93} (for $U_{q}(B^{(1)}_{r})$ or $Y(B_{r})$ ): for $s \in \mathbb{Z}_{\ge 1}$, 
\begin{align}
& \Tb_{a,s}^{[-1]}\Tb_{a,s}^{[1]}=\Tb_{a,s+1}\Tb_{a,s-1}+\Tb_{a-1,s}\Tb_{a+1,s} 
\quad \text{for} \quad 1 \le a \le r-2, 
\label{Tsys1}
\\
& \Tb_{r-1,s}^{[-1]}\Tb_{r-1,s}^{[1]}=\Tb_{r-1,s+1}\Tb_{r-1,s-1}+\Tb_{r-2,s}\Tb_{r,2s}, 
\\
& \Tb_{r,2s}^{[-\frac{1}{2}]}\Tb_{r,2s}^{[\frac{1}{2}]}=\Tb_{r,2s+1}\Tb_{r,2s-1}+\Tb_{r-1,s}^{[-\frac{1}{2}]}\Tb_{r-1,s}^{[\frac{1}{2}]},
\\
& \Tb_{r,2s-1}^{[-\frac{1}{2}]}\Tb_{r,2s-1}^{[\frac{1}{2}]}=\Tb_{r,2s}\Tb_{r,2s-2}+\Tb_{r-1,s-1} \Tb_{r-1,s}
\label{Tsys4}
\end{align} 
with the boundary condition:
\begin{align}
\Tb_{a,0}=1 \quad \text{for} \quad  1 \le a \le r, 
\qquad 
\Tb_{0,s}=\prod_{j=1}^{s}\Qb_{\emptyset}^{[2r-s+2j-1]} \Qb_{\emptyset}^{[s-2j]} \quad \text{for} \quad  s \ge 0. 
\label{Tsys-bc}
\end{align}
This corresponds to [eq.\ (4.17), $p=1$ in \cite{KS94-1}]. 
We also have another expression of the solution, which follows from \eqref{soQQb-2}: for $s\in \mathbb{Z}_{ \ge 0}$,
\begin{align}
\Tb_{a,s}& =   
\Phi_{a,s}
\sum_{I\sqcup J=\Bm, \atop |I|=a} 
\frac{ \prod_{i \in I} z_{i}^{s-a+2r-1}
\prod_{i \in I  } (z_{i}+1)}
{\prod_{(i,j) \in I \times J  } (z_{i}-z_{j}) }  
\Qb^{[s+2r-1]}_{I}  
\Qb^{[-s]}_{J, \Fm}  
\quad 
\text{for} \quad  0 \le a  \le r-1,
\label{Tsum1}
\\
\Tb_{r,2s}& =   
\Phi_{r,2s}
\sum_{I\sqcup J=\Bm, \atop |I|=r} 
\frac{ \prod_{i \in I} z_{i}^{s+r-1}
\prod_{i \in I  } (z_{i}+1)}
{\prod_{(i,j) \in I \times J  } (z_{i}-z_{j}) }  
\Qb^{[s+2r-1]}_{I}  
\Qb^{[-s]}_{J,\Fm}  ,
\\
\Tb_{r,2s+1}& =   
\Phi_{r,2s+1}
\prod_{j=1}^{r}(z_{j}^{\frac{1}{2}}+z_{j}^{-\frac{1}{2}})
\sum_{I\sqcup J=\Bm, \atop |I|=r} 
\frac{ \prod_{i \in I} z_{i}^{s+r}}
{\prod_{(i,j) \in I \times J  } (z_{i}-z_{j}) }  
\Qb^{[s+2r-\frac{1}{2}]}_{I}  
\Qb^{[-s-\frac{1}{2}]}_{J}  .
\label{Tsum3}
\end{align} 
In \eqref{Tsum1}-\eqref{Tsum3}, one can use other Q-functions given by \eqref{Q-Q}.
\subsection{Proof of the solutions}
The following  relation \cite{KOS95}
\footnote{In \cite{KOS95}, \eqref{T+T} is introduced for tableau sum expressions of T-functions. Here we would like to 
show that our Wronskian expressions of T-functions 
satisfy \eqref{T+T}.}
 plays a key role in the proof of the T-system. 
\begin{align}
\Ts_{a,1}^{\Bm,\Fm}+\Ts_{2r-a-1,1}^{\Bm,\Fm}=\Tb_{r,1}^{[r-a-\frac{1}{2}]} \Tb_{r,1}^{[-r+a+\frac{1}{2}]} 
\quad \text{for} \quad a \in \mathbb{Z}.
\label{T+T}
\end{align}
The relation \eqref{Tsys4} for $s=0$ is equivalent to \eqref{T+T}.
We will prove the relation \eqref{T+T} only for the case $a \ge 1$ since this is invariant under 
the exchange $a \leftrightarrow 2r-a-1$. 
For $a \ge 2r$, the second term in the left hand side of \eqref{T+T} vanishes and the first term factorizes due to \eqref{soQQb-1}.
\begin{align}
\Ts_{a,1}^{\Bm,\Fm} 
=\prod_{j=1}^{r}(z_{j}^{\frac{1}{2}}+z_{j}^{-\frac{1}{2}})^{2} 
\Qb_{\Fm}^{[2r-a-1]} \Qb_{\Bm}^{[a]} 
=\Tb_{r,1}^{[r-a-\frac{1}{2}]} \Tb_{r,1}^{[-r+a+\frac{1}{2}]} .
\end{align}
This means that the T-function  factorizes at the boundary of $[2r,1]$-hook. 
Next we consider the case $1 \le a \le 2r-1$. 
The following relation follows from \eqref{QQf}.
\begin{multline}
\frac{\Qb_{b,f}}{z_{b}-z_{f}}=
\frac{(z_{b}z^{-1}_{f})^{c}\Qb^{[2c]}_{b,f} \Qb_{\emptyset}}{(z_{b}-z_{f})\Qb_{\emptyset}^{[2c]}}-z^{-1}_{f}
\sum_{k=1}^{c}(z_{b}z^{-1}_{f})^{k-1} 
\frac{\Qb^{[2k-1]}_{b} \Qb^{[2k-1]}_{f} \Qb_{\emptyset}}{\Qb_{\emptyset}^{[2k-2]}\Qb_{\emptyset}^{[2k]}}
\\ \text{for} 
\quad c \in \mathbb{Z}_{\ge 0} ,
\quad b \in \mathfrak{B}, 
\quad f \in \mathfrak{F}.
\label{Qbfext}
\end{multline}
Let us substitute \eqref{Qbfext} (for $c=2r-a$ with \eqref{z-z}) 
into the first column of the Wronskian expression of the T-function. 
\begin{align}
& \Ts_{a,1}^{\Bm,\Fm} (\Psi_{a,1}^{(2r,1)})^{-1}\Ds (\Bm|\Fm) =(-1)^{r}
\begin{vmatrix}  
\left( \frac{\Qb_{b,2r+1}^{[a-2r]}}{z_{b}+1 } 
\right)_{ b\in \Bm } 
&   
 \left(
z_{b }^{j-1} \Qb_{b}^{[2j-1+a-2r]}
\right)_{  
\genfrac{}{}{0pt}{}{b\in \Bm, }{j \in \langle 1, 2r-a-1 \rangle \sqcup \langle 2r-a+1, 2r \rangle} }       
\end{vmatrix}
\nonumber 
\\
&=(-1)^{r}
\begin{vmatrix}  
\left(
\frac{(-z_{b})^{2r-a} \Qb^{[2r-a]}_{b,2r+1}\Qb_{\emptyset}^{[a-2r]}}{(z_{b}+1)\Qb_{\emptyset}^{[2r-a]}}
\right)_{ b\in \Bm } 
&   
 \left(
z_{b }^{j-1} \Qb_{b}^{[2j-1+a-2r]}
\right)_{  
\genfrac{}{}{0pt}{}{b\in \Bm, }{j \in \langle 1, 2r-a-1 \rangle \sqcup \langle 2r-a+1, 2r \rangle} }       
\end{vmatrix}
\nonumber 
\\
& \qquad +(-1)^{r}
\sum_{k=1}^{2r-a}
\Biggl|
\left(
(-z_{b})^{k-1} \frac{\Qb^{[2k-1+a-2r]}_{b} \Qb^{[2k-1+a-2r]}_{2r+1} \Qb_{\emptyset}^{[a-2r]}}
{\Qb_{\emptyset}^{[2k-2+a-2r]} \Qb_{\emptyset}^{[2k+a-2r]}} 
\right)_{ b\in \Bm } 
\nonumber 
\\
&   
\hspace{160pt}
 \left(
z_{b }^{j-1} \Qb_{b}^{[2j-1+a-2r]}
\right)_{  
\genfrac{}{}{0pt}{}{b\in \Bm, }{j \in \langle 1, 2r-a-1 \rangle \sqcup \langle 2r-a+1, 2r \rangle} }
\Biggr| .
\end{align}
The summation over $k \in \{1,2,\dots 2r-a-1\}$ vanishes because of proportionality of  two columns of matrices.
\begin{align}
& \Ts_{a,1}^{\Bm,\Fm}\Ds (\Bm|\Fm) 
=(-1)^{a+r} A
\begin{vmatrix}  
\left(
\frac{z_{b}^{2r-a} \Qb^{[2r-a]}_{b,2r+1}}{z_{b}+1}
\right)_{ b\in \Bm } 
&   
 \left(
z_{b }^{j-1} 
\Qb_{b}^{[2j-1+a-2r]}
\right)_{  
\genfrac{}{}{0pt}{}{b\in \Bm, }{j \in \langle 1, 2r-a-1 \rangle \sqcup \langle 2r-a+1, 2r \rangle} }       
\end{vmatrix}
\nonumber 
\\
& \qquad +(-1)^{r}
\frac{\Psi_{a,1}^{(2r,1)} \Qb_{\emptyset}^{[a-2r]} \Qb^{[2r-1-a]}_{2r+1} }{\Qb_{\emptyset}^{[2r-a-2]}\Qb_{\emptyset}^{[2r-a]}}
\begin{vmatrix}
 \left(
z_{b }^{j-1} \Qb_{b}^{[2j-1+a-2r]}
\right)_{  
\genfrac{}{}{0pt}{}{b\in \Bm, }{j \in \langle 1, 2r \rangle} }
\end{vmatrix}
\nonumber 
\\
&=(-1)^{a+r} A
\begin{vmatrix}  
\left(
\frac{ \Qb^{[2r-a]}_{b,2r+1}}{z_{b}+1}
\right)_{ b\in \Bm } 
&   
 \left(
z_{b }^{j-1-2r+a} \Qb_{b}^{[2j-1+a-2r]}
\right)_{  
\genfrac{}{}{0pt}{}{b\in \Bm, }{j \in \langle 1, 2r-a-1 \rangle \sqcup \langle 2r-a+1, 2r \rangle} }       
\end{vmatrix}
\nonumber 
\\
& \qquad +\Ds (\Bm|\emptyset) 
\Qb^{[2r-1-a]}_{2r+1}
\Qb^{[a]}_{\Bm} 
\qquad \qquad \qquad \text{[cf. \eqref{emptyT}, \eqref{proz=1}]}
\nonumber 
\\
&=(-1)^{r+1} A
\begin{vmatrix}  
 \left(
z_{b }^{j-1-2r+a} \Qb_{b}^{[2j-1+a-2r]}
\right)_{  
\genfrac{}{}{0pt}{}{b\in \Bm, }{j \in \langle 1, 2r-a-1 \rangle } }   
&
\left(
\frac{ \Qb^{[2r-a]}_{b,2r+1}}{z_{b}+1}
\right)_{ b\in \Bm } 
&   
 \left(
z_{b }^{j-1} \Qb_{b}^{[2j-1-a+2r]}
\right)_{  
\genfrac{}{}{0pt}{}{b\in \Bm, }{j \in \langle 1, a \rangle} }       
\end{vmatrix}
\nonumber 
\\
& \qquad +\Ds (\Bm|\Fm) 
\Tb_{r,1}^{[r-a-\frac{1}{2}]} \Tb_{r,1}^{[-r+a+\frac{1}{2}]} 
\qquad \qquad \text{[cf. \eqref{Tr1}]} ,
 \label{TaD}
\end{align} 
where $A=\Psi_{a,1}^{(2r,1)} \Qb_{\emptyset}^{[a-2r]} (\Qb_{\emptyset}^{[2r-a]})^{-1}$, 
 and the following relation is used in the last step:
\begin{align}
\frac{\Ds (\Bm|\emptyset)}{\Ds (\Bm|\Fm)}=\prod_{j=1}^{r}(z_{j}^{\frac{1}{2}}+z_{j}^{-\frac{1}{2}})^{2} .
\label{D-D}
\end{align}
Let us apply the Laplace expansion formula to the first term in the right hand side of \eqref{TaD}.
\begin{align}
&
\begin{vmatrix}  
 \left(
z_{b }^{j-1-2r+a} \Qb_{b}^{[2j-1+a-2r]}
\right)_{  
\genfrac{}{}{0pt}{}{b\in \Bm, }{j \in \langle 1, 2r-a-1 \rangle } }   
&
\left(
\frac{ \Qb^{[2r-a]}_{b,2r+1}}{z_{b}+1}
\right)_{ b\in \Bm } 
&   
 \left(
z_{b }^{j-1} \Qb_{b}^{[2j-1-a+2r]}
\right)_{  
\genfrac{}{}{0pt}{}{b\in \Bm, }{j \in \langle 1, a \rangle} }       
\end{vmatrix}
\nonumber 
\\
&=\sum_{I \sqcup J= \Bm \atop |I|=2r-a-1} \epsilon_{I; J} 
\left(\prod_{b \in I}z_{b}^{a-2r} \right)
\Delta^{I,\emptyset,[a-2r ]}_  {\emptyset, \langle 1, 2r-a-1 \rangle} 
\Delta^{J,\emptyset,[2r-a ]}_  {\Fm, \langle 1, a \rangle} 
\nonumber 
\\
&= B \sum_{I \sqcup J= \Bm \atop |I|=2r-a-1} \epsilon_{I; J} (-1)^{\frac{(2r-a-1)(2r-a-2)}{2}+\frac{(a+1)a}{2}}
\left(\prod_{b \in I}z_{b}^{a-2r} \right)
\Ds (I|\emptyset)  \Ds (J|\Fm) 
\Qb_{I}^{[-1]}  \Qb_{J,\Fm}^{[2r]}
\nonumber \\
& \hspace{200pt} \text{[by \eqref{emptyT}]} 
\nonumber 
\\
&=B \sum_{I \sqcup J= \Bm \atop |I|=2r-a-1} \epsilon_{I; J} (-1)^{\frac{(2r-a-1)(2r-a-2)}{2}+\frac{(a+1)a}{2}}
\left(\prod_{b \in I}z_{b}^{a-2r} \right)
\Ds (I|\emptyset)  \Ds (J|\Fm) 
\Qb_{J^{*},\Fm}^{[-1]}  \Qb_{I^{*}}^{[2r]}
\nonumber \\
& \hspace{200pt} \text{[by \eqref{Q-Q}]} 
\nonumber 
\\
&=(-1)^{r}B\,  \Ds (\Bm|\Fm)  \sum_{I \sqcup J= \Bm \atop |I|=2r-a-1} 
\frac{\prod_{b \in I^{*}}z_{b}^{a+1}(z_{b}+1)}{\prod_{(b,b^{\prime}) \in I^{*} \times J^{*}} (z_{b}-z_{b^{\prime}})}
\Qb_{I^{*}}^{[2r]} \Qb_{J^{*},\Fm}^{[-1]}  
\nonumber 
\\
&=(-1)^{r} B\, \Ds (\Bm|\Fm) \Ts^{\Bm,\Fm}_{2r-a-1,1} 
\hspace{76pt} \text{[by \eqref{soQQb-2}]} 
\label{TaD2}
\end{align}
where $\epsilon_{I;J}=(-1)^{\mathrm{Card}\{ (i,j) \in I \times J| i>j \}}$, 
$B= \Qb_{\emptyset}^{[2r-a-2]} \Qb_{\emptyset}^{[a+2r]} (\Psi_{\emptyset}^{(2r-a-1,0)[-1]} \Psi_{\emptyset}^{(a+1,1)[2r]} )^{-1}$, 
and the following relation is used:
\begin{align}
\frac{\Ds (I|\emptyset)\Ds (J|\Fm)}{\Ds (\Bm|\Fm)}=(-1)^{a+1}\epsilon_{I;J}
\frac{ \prod_{b \in I^{*}}z_{b}^{2a-2r+1}(z_{b}+1)}{\prod_{(b,b^{\prime}) \in I^{*} \times J^{*}}(z_{b}-z_{b^{\prime}})} .
\label{D-D2}
\end{align}
In the last step of \eqref{TaD2}, we have used $I \sqcup J=I^{*} \sqcup J^{*}= \Bm$, $|I|=|I^{*}|$. 
Substituting \eqref{TaD2} into \eqref{TaD} with the relation $AB=1$, we arrive at \eqref{T+T}.
Now we would like to prove the following CBR type formula:
\begin{align}
\Tb_{a,s}&=
\begin{vmatrix}
\left(\Ts_{a-i+j,1}^{\Bm, \Fm \, [s-i-j+1]}\right)_{1 \le i \le s \atop 1 \le j \le s} 
\end{vmatrix}
\qquad \text{for} \quad 0 \le a \le r-1
,
\label{CBRten}
\\
\Tb_{r,2s}&=
\begin{vmatrix}
\left(\Ts_{r-i+j,1}^{\Bm, \Fm \, [s-i-j+1]}\right)_{1 \le i \le s \atop 1 \le j \le s} 
\end{vmatrix}
,
\label{CBRse}
\\
\Tb_{r,2s+1}&=
\begin{vmatrix}
\left(\Ts_{r-i+j,1}^{\Bm, \Fm \, [s-i-j+\frac{3}{2}]}\right)_{1 \le i \le s+1 \atop 1 \le j \le s} 
&
\left(\Tb_{r,1}^{ [s-2i+2]}\right)_{1 \le i \le s+1 } 
\end{vmatrix},
\label{CBRso}
\end{align}
where $s \in {\mathbb Z}_{\ge 0}$. 
It is known [Theorem 5.1 in \cite{KOS95}] that these determinants \eqref{CBRten}-\eqref{CBRso} solve the T-system 
\eqref{Tsys1}-\eqref{Tsys4} under the condition that the matrix elements satisfy the functional relation \eqref{T+T}.
 The relation \eqref{CBRso} for $s=0$ is trivial. Then we assume  $s \ge 1$. 
Let us substitute \eqref{Trec0} for $(a,s)=(r-i+j,1)$ (with \eqref{T=F}, \eqref{TF-rel03}) and \eqref{Tr1} into the right hand side of \eqref{CBRso}, and 
sort out the last column by row operations (except for the last row):
\begin{multline}
[\text{r.h.s. of \eqref{CBRso}}]=
\\
=
\begin{vmatrix}
\left( \Qb_{\Bm,\Fm}^{[r+s-2i+\frac{1}{2}]} \Qb_{\emptyset}^{[r+s-2j+\frac{3}{2}]}
{\mathcal F}_{(1^{r-i+j})}^{\Bm, \Fm \, [s-i-j+\frac{3}{2}]}\right)_{1 \le i \le s+1 \atop 1 \le j \le s} 
&
\left(
\prod_{k=1}^{r}(z_{k}^{\frac{1}{2}}+z_{k}^{-\frac{1}{2}})
\Qb_{\Bm}^{ [s-2i+r+\frac{3}{2}]}\right)_{1 \le i \le s+1 } 
\end{vmatrix}
\\
= 
\prod_{j=1}^{s}\Qb_{\emptyset}^{[r+s-2j+\frac{3}{2}]}
\prod_{k=1}^{r}(z_{k}^{\frac{1}{2}}+z_{k}^{-\frac{1}{2}}) 
\begin{vmatrix}
\left( A_{ij}
\right)_{1 \le i \le s+1 \atop 1 \le j \le s} 
&
\left(
\delta_{i,s+1} \Qb_{\Bm}^{ [s-2i+r+\frac{3}{2}]}\right)_{1 \le i \le s+1 } 
\end{vmatrix}
\\
=
\prod_{j=1}^{s}\Qb_{\emptyset}^{[r+s-2j+\frac{3}{2}]}
\prod_{k=1}^{r}(z_{k}^{\frac{1}{2}}+z_{k}^{-\frac{1}{2}})
\Qb_{\Bm}^{ [-s+r-\frac{1}{2}]} 
\\
\times 
\begin{vmatrix}
\left( \Qb_{\Bm,\Fm}^{[r+s-2i+\frac{1}{2}]}  {\mathcal F}_{(1^{r-i+j})}^{\Bm, \Fm \, [s-i-j+\frac{3}{2}]}
-\frac{ \Qb_{\Bm}^{ [s-2i+r+\frac{3}{2}]} \Qb_{\Bm,\Fm}^{[r+s-2i-\frac{3}{2}]} }
{\Qb_{\Bm}^{ [s-2i+r-\frac{1}{2}]}}
{\mathcal F}_{(1^{r-i+j-1})}^{\Bm, \Fm \, [s-i-j+\frac{1}{2}]}
\right)_{1 \le i \le s \atop 1 \le j \le s} 
\end{vmatrix}
, \label{CBRso2}
\end{multline}
where $\theta(\text{True})=1$, $\theta(\text{False})=0$, and 
\begin{align*}
A_{ij}=\Qb_{\Bm,\Fm}^{[r+s-2i+\frac{1}{2}]} 
{\mathcal F}_{(1^{r-i+j})}^{\Bm, \Fm \, [s-i-j+\frac{3}{2}]}
-\frac{ \theta(1 \le i \le s)\Qb_{\Bm}^{ [s-2i+r+\frac{3}{2}]}  \Qb_{\Bm,\Fm}^{[r+s-2i-\frac{3}{2}]} }
{\Qb_{\Bm}^{ [s-2i+r-\frac{1}{2}]}}
{\mathcal F}_{(1^{r-i+j-1})}^{\Bm, \Fm \, [s-i-j+\frac{1}{2}]} .
\end{align*}
Let $I_{a}=(1,2,\dots, a)$ for $1 \le a \le 2r+1$. 
For $a \in \mathbb{Z}_{\ge 0}$, we have
\begin{align}
{\mathcal  F}_{(1^{a})}^{I_{2r+1}}&=
\sum_{t\in\mathsf{Tab}_{I_{2r+1}}((1^{a}))}
\prod_{j=1}^{a}
p_{t_{j,1}}
{\mathcal  X}_{I_{t_{j,1}}}^{[-a+2j-2+2r]}
\nonumber \\
&=
\sum_{k=0}^{\min\{a,2r\}}
(-1)^{a-k}
\sum_{t\in\mathsf{Tab}_{I_{2r}}((1^{k}))}
\prod_{j=1}^{k}
{\mathcal  X}_{I_{t_{j,1}}}^{[-a+2j-2+2r]}
\prod_{j=k+1}^{a}
{\mathcal  X}_{I_{2r+1}}^{[-a+2j-2+2r]}
\nonumber \\
&=
\sum_{k=0}^{\min\{a,2r\}}
\frac{{\mathcal F}_{(1^{k})}^{I_{2r}\, [k-a-1]} \Qb_{I_{2r}}^{[a]} \Qb_{I_{2r+1}}^{[2k-a-1]} }{\Qb_{I_{2r}}^{[2k-a]} \Qb_{I_{2r+1}}^{[a-1]} } ,
\label{F=sum} 
\end{align}
where we use  \eqref{boxes} and \eqref{DVF-tab1}. 
This means that the relation 
\begin{align}
{\mathcal  F}_{(1^{a})}^{\Bm, \Fm}
&=
\sum_{k=0}^{\min\{a,2r\}}
\frac{{\mathcal F}_{(1^{k})}^{\Bm , \emptyset  \, [k-a-1]} \Qb_{\Bm}^{[a]}  \Qb_{\Bm, \Fm}^{[2k-a-1]} }
{\Qb_{\Bm}^{[2k-a]}  \Qb_{\Bm, \Fm}^{[a-1]}} 
\label{F=sum2} 
\end{align}
holds. 
Substituting \eqref{F=sum2} into the right hand side of \eqref{CBRso2}, we obtain
\begin{align}
& [\text{r.h.s. of \eqref{CBRso2}}]= \nonumber
\\
&=
\prod_{j=1}^{s}\Qb_{\emptyset}^{[r+s-2j+\frac{3}{2}]}
\prod_{k=1}^{r}(z_{k}^{\frac{1}{2}}+z_{k}^{-\frac{1}{2}})
\Qb_{\Bm}^{ [-s+r-\frac{1}{2}]} 
\begin{vmatrix}
\left(
\Qb_{\Bm,\Fm}^{ [s+r-2i+\frac{1}{2}]} 
 {\mathcal F}_{(1^{r-i+j})}^{\Bm, \emptyset \, [s-i-j+\frac{1}{2}]}
\right)_{1 \le i \le s \atop 1 \le j \le s} 
\end{vmatrix}
\nonumber \\
&=
\prod_{j=1}^{s}\Qb_{\emptyset}^{[r+s-2j+\frac{3}{2}]}
\Qb_{\Bm,\Fm}^{ [s+r-2j+\frac{1}{2}]} 
\prod_{k=1}^{r}(z_{k}^{\frac{1}{2}}+z_{k}^{-\frac{1}{2}})
\Qb_{\Bm}^{ [-s+r-\frac{1}{2}]}
{\mathcal F}_{(s^{r})}^{\Bm, \emptyset \, [-\frac{1}{2}]} 
\qquad \text{[by \eqref{superJT1}]} 
\nonumber \\
&=\Phi_{r,2s+1}
\prod_{k=1}^{r}(z_{k}^{\frac{1}{2}}+z_{k}^{-\frac{1}{2}})
{\mathsf F}_{(s^{r})}^{\Bm, \emptyset \, [-\frac{1}{2}]}
\qquad \text{[by \eqref{TF-rel03}]}
\nonumber \\
&=\Phi_{r,2s+1}
\prod_{k=1}^{r}(z_{k}^{\frac{1}{2}}+z_{k}^{-\frac{1}{2}})
\Ts_{(s^{r})}^{\Bm, \emptyset \, [-\frac{1}{2}]}
\qquad \text{[by \eqref{T=F}]}
. \label{CBRso3}
\end{align}
Combining \eqref{Trec0}, \eqref{solB3}, \eqref{CBRso2} and \eqref{CBRso3}, we arrive at \eqref{CBRso}. 
The proofs of \eqref{CBRten} and \eqref{CBRse} are similar (but easier) to the proof of \eqref{CBRso} 
[use  \eqref{superJT1}, \eqref{TF-rel03}, \eqref{Trec0}, \eqref{T=F}, \eqref{solB1}, \eqref{solB2}].
\subsection{Another expression of the T-function for the spinorial representation} 
There is an expression of the T-function for  the spinorial representation, which is written in terms of ratios of Q-functions. 
Our claim is that it coincides with the Wronskian formula \eqref{Tr1}:
\begin{align}
\Tb_{r,1}=\sum_{s_{1},\dots , s_{r}=\pm 1}
\Qb_{\emptyset}^{[-s_{1}+r-\frac{1}{2}]}
\prod_{a=1}^{r-1} 
z_{a}^{\frac{s_{a}}{2}}
\left(
\frac{\Qb_{1,2\dots , a}^{[r+\frac{1}{2}+\frac{\rho_{a}}{2}]}}{\Qb_{1,2\dots , a}^{[r-\frac{3}{2}+\frac{\rho_{a}}{2}]}}
\right)^{\frac{s_{a}-s_{a+1}}{2}}
z_{r}^{\frac{s_{r}}{2}}
\left(
\frac{\Qb_{1,2\dots , r}^{[r+\frac{\rho_{r}}{2}]}}{\Qb_{1,2\dots , r}^{[r-1+\frac{\rho_{r}}{2}]}}
\right)^{s_{r}},
\label{Tspf}
\end{align}
where $\rho_{a}=2(s_{1}+s_{2}+\dots +s_{a-1})+(s_{a}-s_{a+1})/(1+\delta_{a,r})$, $s_{r+1}=-s_{r}$. 
The right hand side of \eqref{Tspf} can be found in  [eq.(7.15) in \cite{KNS10}] (see also \cite{Re83,KS94-1,KOS95}).
\footnote{Here we inserted the boundary twist parameters $\{z_{a}\}$ and the function $\Qb_{\emptyset}$, 
which are not present in \cite{KNS10}.} 
The relation \eqref{Tspf} is deduced if one admits that both l.h.s. and r.h.s. of \eqref{Tspf} is a part of a solution of the T-system.
The former statement is proven in the previous subsection. The latter is equivalent to the statement that 
[eq. (7.15) in \cite{KNS10}] coincides with [eq. (2.10) for $a=r$ in \cite{KOS95}].  
A direct proof of \eqref{Tspf} for $r=2$ is given as follows.  First, we rewrite \eqref{Tspf} for $r=2$ 
as
\begin{multline}
\Qb_{1}^{[\frac{3}{2}]} \Qb_{12}^{[\frac{1}{2}]} \Qb_{12}^{[\frac{5}{2}]} \Tb_{2,1}=
z_{1}^{\frac{1}{2}} z_{2}^{\frac{1}{2}}
\Qb_{\emptyset}^{[\frac{1}{2}]}
\Qb_{1}^{[\frac{3}{2}]} \Qb_{12}^{[\frac{1}{2}]} \Qb_{12}^{[\frac{7}{2}]}
+
z_{1}^{\frac{1}{2}} z_{2}^{-\frac{1}{2}}
\Qb_{\emptyset}^{[\frac{1}{2}]}
\Qb_{1}^{[\frac{7}{2}]} \Qb_{12}^{[\frac{1}{2}]} \Qb_{12}^{[\frac{3}{2}]}
\\
+
z_{1}^{-\frac{1}{2}} z_{2}^{\frac{1}{2}}
\Qb_{\emptyset}^{[\frac{5}{2}]}
\Qb_{1}^{[-\frac{1}{2}]} \Qb_{12}^{[\frac{3}{2}]} \Qb_{12}^{[\frac{5}{2}]}
+
z_{1}^{-\frac{1}{2}} z_{2}^{-\frac{1}{2}}
\Qb_{\emptyset}^{[\frac{5}{2}]}
\Qb_{1}^{[\frac{3}{2}]} \Qb_{12}^{[-\frac{1}{2}]} \Qb_{12}^{[\frac{5}{2}]} .
\label{Tspf2}
\end{multline}
We can check this step by step: 
\begin{align}
& [\text{l.h.s. of \eqref{Tspf2}}]=
\prod_{j=1}^{2}(z_{j}^{\frac{1}{2}}+z_{j}^{-\frac{1}{2}})
\underbrace{\Qb_{1}^{[\frac{3}{2}]}\Qb_{5}^{[\frac{3}{2}]}}_{\text{apply \eqref{QQf}}}
 \Qb_{12}^{[\frac{1}{2}]} 
 \Qb_{12}^{[\frac{5}{2}]} 
\qquad \qquad \text{[by \eqref{Tr1}]} 
\nonumber
\\
&=(z_{2}^{\frac{1}{2}}+z_{2}^{-\frac{1}{2}})
\bigl(
z_{1}^{\frac{1}{2}} \Qb_{\emptyset}^{[\frac{1}{2}]}  \Qb_{12}^{[\frac{1}{2}]} 
\underbrace{\Qb_{12}^{[\frac{5}{2}]} \Qb_{15}^{[\frac{5}{2}]}}_{\text{apply \eqref{QQf}}}
+ z_{1}^{-\frac{1}{2}}
\underbrace{ \Qb_{12}^{[\frac{1}{2}]} \Qb_{15}^{[\frac{1}{2}]}}_{\text{apply \eqref{QQf}}}
 \Qb_{\emptyset}^{[\frac{5}{2}]}  \Qb_{12}^{[\frac{5}{2}]}
\bigr)
\nonumber
\\
&= z_{1}^{\frac{1}{2}} z_{2}^{-\frac{1}{2}} \Qb_{\emptyset}^{[\frac{1}{2}]} \Qb_{12}^{[\frac{1}{2}]} 
\bigl( z_{2}\Qb_{1}^{[\frac{3}{2}]}\Qb_{125}^{[\frac{7}{2}]} +\Qb_{1}^{[\frac{7}{2}]}\Qb_{125}^{[\frac{3}{2}]}   \bigr)
+
z_{1}^{-\frac{1}{2}} z_{2}^{-\frac{1}{2}} \Qb_{\emptyset}^{[\frac{5}{2}]}
\bigl( z_{2}\Qb_{1}^{[-\frac{1}{2}]}\Qb_{125}^{[\frac{3}{2}]} +\Qb_{1}^{[\frac{3}{2}]}\Qb_{125}^{[-\frac{1}{2}]}   \bigr)
\Qb_{12}^{[\frac{5}{2}]} 
\nonumber
\\
&=[\text{r.h.s. of \eqref{Tspf2}}] 
\qquad \qquad \text{[by \eqref{Q-Q}]} ,
\end{align}
where \eqref{z-z}, namely, $z_{5}=-1$ is used. 
We expect that \eqref{Tspf} for general $r$ can be proven by applying similar procedures recursively. 
 
\subsection{Wronskian formulas for non-rectangular Young diagrams}
There are CBR or tableau sum expressions of T-functions for non-rectangular Young diagrams \cite{KOS95}. 
In this section, we propose Wronskian-type expressions of them. 
For any skew-Young diagram $\lambda \subset \mu$, we introduce the following CBR-type determinants:
\begin{align}
\Tb_{\lambda \subset \mu}&= 
\begin{vmatrix}
  \left(\Ts^{\Bm,\Fm \, [\mu_{1}-\mu_{1}^{\prime}+\mu_{i}^{\prime}+\lambda_{j}^{\prime}-i-j+1]}_{\mu_{i}^{\prime}-\lambda_{j}^{\prime}-i+j ,1}
   \right)_{1 \le i \le \mu_{1} \atop
   1 \le j \le \mu_{1}}
  \end{vmatrix} ,
 \label{JT-wron}
\end{align}
\begin{align}
{\mathbb S}_{\lambda \subset \mu}&=
\begin{vmatrix}
\left(\Ts_{\mu_{i}^{\prime}-\lambda_{j}^{\prime}-i+j,1}^{\Bm, \Fm \, [2\mu_{1}+\mu_{i}^{\prime}+\lambda_{j}^{\prime}-i-j-r-\frac{1}{2}]}\right)_{1 \le i \le \mu_{1} \atop 1 \le j \le \mu_{1}-1} 
&
\left(\Tb_{r,1}^{ [2\mu_{1}+2\mu_{i}^{\prime}-2i-2r]}\right)_{1 \le i \le \mu_{1} } 
\end{vmatrix}
.
\label{CBRsogen}
\end{align}
The first equation \eqref{JT-wron} corresponds
\footnote{In \cite{KOS95}, \eqref{JT-wron} and \eqref{CBRsogen} are introduced for tableau sum expressions of T-functions. 
In addition, Drinfeld polynomials, which characterize representation theoretical meaning of these, are also available in \cite{KOS95}. }
to eq.\ (3.5a) in \cite{KOS95}, which is related to tensor representations (spin even representations).
The second equation \eqref{CBRsogen} corresponds
to eq.\ (4.8a) in \cite{KOS95}, which is related to spin odd representations.
Our claim is that these are expressed as Wronskian-type determinants: 
\begin{align}
\Tb_{\tilde{\mu}}&=\Phi_{\mu} \Ts^{\Bm,\Fm}_{\mu}
 ,
\label{speven}
\\
{\mathbb S}_{\widetilde{\mu +(1^{r})}}^{[-\mu_{1}+\mu_{1}^{\prime}-r]}&=
\Qb_{\Bm,\Fm}^{[-\mu_{1}+\mu_{1}^{\prime}+\frac{1}{2}]} \Phi_{\mu}^{[\frac{1}{2}]} 
\prod_{j=1}^{r}(z_{j}^{\frac{1}{2}}+z_{j}^{-\frac{1}{2}})\Ts^{\Bm,\emptyset \, [-\frac{1}{2}]}_{\mu}
 ,
\label{spinodd}
\end{align}
where 
$\mu=(\mu_{1},\mu_{2}, \dots , \mu_{r})$, $\mu_{1} \ge \mu_{2} \ge \dots \ge \mu_{r} \ge 0$, 
$\widetilde{\mu +(1^{r})}= (\mu_{1}-\mu_{r},\mu_{1}-\mu_{r-1}, \dots , \mu_{1}-\mu_{2},0) \subset ((\mu_{1}+1)^{r})$, 
$(\mu_{1}-\mu_{r},\mu_{1}-\mu_{r-1}, \dots , \mu_{1}-\mu_{2},0)^{\prime}
=(r-\mu_{\mu_{1}}^{\prime} , r-\mu_{\mu_{1}-1}^{\prime}, \dots , r-\mu_{1}^{\prime})$, 
and for \eqref{spinodd}, $\mu_{1}^{\prime} \le r$. 
The normalization function is defined by
\begin{align}
\Phi_{\mu}=\frac{\prod_{j=1}^{\mu_{1}} 
\Qb_{\emptyset}^{[2r-1-\mu_{1}+\mu_{1}^{\prime}-2\mu_{j}^{\prime}+2j]}\Qb_{\Bm,\Fm}^{[\mu_{1}+\mu_{1}^{\prime}-2j]} }
{\Qb_{\emptyset}^{[2r-1+\mu_{1}-\mu_{1}^{\prime}]}\Qb_{\Bm,\Fm}^{[-\mu_{1}+\mu_{1}^{\prime}]}}  .
\end{align}
The expression \eqref{spinodd} for $\mu=(s^{r})$ corresponds to \eqref{solB3}.  
The proof of \eqref{spinodd} is parallel to that of \eqref{CBRso}
 (use  \eqref{superJT1}, \eqref{TF-rel03}, \eqref{T=F},  \eqref{Tr1}, \eqref{F=sum2}). 
As for \eqref{speven}, it is proven only for rectangular Young diagrams (because of \eqref{T=F} for $|B||F| \ne 0$). 
The right hand side of \eqref{speven} and \eqref{spinodd} can be expressed as
 summations over (quotients of) symmetric groups (see Proposition 3.6 in \cite{T09}). 

T-functions for typical representations factorizes. 
Based on properties of the Wronskian determinant \eqref{unnor-t1}, 
we can show [cf. eq. (4.5) in \cite{T09}]:
\begin{align}
\Ts^{\Bm,\Fm \, [\mu_{1}^{\prime}-r-c-\frac{1}{2}]}_{\mu +(1^{r+c})}&=
\left(
\frac{\Qb_{\emptyset}^{[\mu_{1}^{\prime}+\mu_{1}-2c-\frac{1}{2}]}}
{\Qb_{\emptyset}^{[\mu_{1}^{\prime}-\mu_{1}-2c-\frac{1}{2}]} \Qb_{\emptyset}^{[-\mu_{1}^{\prime}+\mu_{1}+2r-\frac{1}{2}]}}
\right)
\prod_{j=1}^{r}(z_{j}^{\frac{1}{2}}+z_{j}^{-\frac{1}{2}})^{2} 
\Qb_{2r+1}^{[\mu_{1}^{\prime}-\mu_{1}-2c-\frac{3}{2}]} \Ts^{\Bm,\emptyset \, [-\frac{1}{2}]}_{\mu},
 \label{facT}
\end{align}
where $c \in \mathbb{Z}_{> r}$, $\mu_{1}^{\prime} \le r$. Thus the T-function factories with respect to 
T-functions for spinorial representations (\eqref{Tr1} and \eqref{spinodd}). 
In order to remove one of them, we would like to take a limit with respect to $c$. 
We find that  the right hand side of \eqref{spinodd} is, up to an overall factor, an asymptotic limit of  \eqref{facT}.
\begin{align}
\lim_{c}
 \Qb_{\emptyset}^{[-\mu_{1}^{\prime}+\mu_{1}+2r-\frac{1}{2}]}
\prod_{j=1}^{r}(z_{j}^{\frac{1}{2}}+z_{j}^{-\frac{1}{2}})^{-1}
\Ts^{\Bm,\Fm \, [\mu_{1}^{\prime}-r-c-\frac{1}{2}]}_{\mu +(1^{r+c})}=
\prod_{j=1}^{r}(z_{j}^{\frac{1}{2}}+z_{j}^{-\frac{1}{2}}) 
 \Ts^{\Bm,\emptyset \, [-\frac{1}{2}]}_{\mu},
\end{align}
where we assume
\begin{align}
\lim_{c} \Qb_{\emptyset}^{[-2c]} =\lim_{c} \Qb_{2r+1}^{[-2c]}=1 . \label{Qlim}
\end{align}
The limit
\footnote{As an example, let us consider Q-functions for a spin chain (the quantum space is $L$-fold tensor product of the fundamental representation). 
Suppose the Q-functions have the form
 $\Qb_{I}=\prod_{j=1}^{n_{I}}(1-u/u_{j}^{I})$, where $u \in {\mathbb C}$, $n_{I} \in {\mathbb Z}_{\ge 0}$, $I \subset {\mathfrak I}$; 
$u_{j}^{I} \in {\mathbb C}$ for $I \ne \emptyset,  {\mathfrak I}$ are Bethe roots; 
$n_{\emptyset}=n_{{\mathfrak I}}=L$ (the number of the lattice site);
 $u^{\emptyset}_{j}=u^{{\mathfrak I}}_{j}=w_{j} \in {\mathbb C}$ (inhomogeneity on the spectral parameter on the $j$-th site in the quantum space). 
 Then in the limit $q^{-c} \to 0$, we have 
$\lim_{c}\Qb_{I}^{[-c]}=\lim_{c}\prod_{j=1}^{n_{I}}(1-uq^{-c}/u_{j}^{I})=1$, where the multiplicative shift with the unit $\hbar =1$ 
is adopted. 
In this normalization of the Q-functions, the parameters $\{z_{j}\}$ depend on the parameters $\{n_{I}\}$.}
 is intended to be $c \to \infty $, but other limits (for example, $c \to -\infty$), which realize \eqref{Qlim}, 
may be necessary depending on the function form of the Q-functions. 

Finally we remark that there are Wronskian-type formulas of T-functions 
$\overline{\mathsf T}_{\mu}^{B,F}$  in terms of the Q-functions $\overline{\Qb}_{I}=\Qb_{\mathfrak{I} \setminus I}$, 
$I \subset \mathfrak{I}$ [cf. eq. (3.14) in \cite{T09}]. The discussions for $\overline{\mathsf T}_{\mu}^{B,F}$ 
are parallel to those for ${\mathsf T}_{\mu}^{B,F}$, but we do not need 180 degree rotation of Young diagrams 
[cf. eqs. (2.45), (2.46), (3.67), (3.68) in \cite{T09}].  We may need [eq. (4.6a), \cite{KOS95}] instead of [eq. (4.8a), \cite{KOS95}] 
for spinorial representations.
 \subsection{Character limit: solution of the Q-system}
The T-functions reduce to characters of  representations of  $U_{q}(B_{r}^{(1)})$ or $Y(B_{r})$  in the limit (or formal replacement) 
 $\Qb_{I} \to 1$ for all $I \subset \Bm \sqcup \Fm$. 
 In particular, the T-functions $\Tb_{a,s}$  reduce to the characters $\chi_{a,s}$ of Kirillov-Reshetikhin modules, 
 which are solutions of the Q-system \cite{KR90} defined, for $s \in {\mathbb Z}_{\ge 1}$,  by 
 \begin{align}
& (\chi_{a,s})^{2}=\chi_{a,s+1}\chi_{a,s-1}+\chi_{a-1,s}\chi_{a+1,s} 
\quad \text{for} \quad 1 \le a \le r-2, 
\label{Qsys1}
\\
&( \chi_{r-1,s})^{2}=\chi_{r-1,s+1}\chi_{r-1,s-1}+\chi_{r-2,s}\chi_{r,2s}, 
\\
& (\chi_{r,2s})^{2}=\chi_{r,2s+1}\chi_{r,2s-1}+(\chi_{r-1,s})^{2} ,
\\
& (\chi_{r,2s-1})^{2}=\chi_{r,2s}\chi_{r,2s-2}+\chi_{r-1,s-1} \chi_{r-1,s}
\label{Qsys4}
\end{align} 
with the boundary condition.
\begin{align}
\chi_{a,0}=1 \quad \text{for} \quad  1 \le a \le r, 
\quad 
\chi_{0,s}=1 \quad \text{for} \quad  s \ge 0. 
\end{align}
The formulas \eqref{speven} and \eqref{spinodd} reduce to Weyl-type formulas of characters 
beyond  Kirillov-Reshetikhin modules. 
\section{Concluding remarks}
Our goals are three fold: 
(i) to establish Wronskian-type formulas of T-functions (analogues of the Weyl character formula)
 associated with any quantum affine (super)algebras or Yangians, 
(ii) to realize them as operators (through Baxter Q-operators), 
(iii) to understand their connection to soliton theory. 
In this paper, we focused on (i) and (iii), and 
explained how to obtain a Wronskian solution of the T-system (discrete Toda field equation) associated with $U_{q}(B_{r}^{(1)})$ 
as a reduction of the Wronskian solution associated with  $U_{q}(gl(2r|1)^{(1)})$. 
It is an interesting problem to consider different types of reductions. 
In particular, the case $(M,N)=(2r,2s+1)$ is related to quantum integrable systems associated with 
$U_{q}(osp(2r+1|2s)^{(1)})$. 
In the paper \cite{T99}, we considered tableaux sum or CBR-type expressions 
of T-functions only for tensor representations of $Y(osp(r|2s))$. 
It is desirable to extend the analysis to spinorial representations and establish 
Wronskian type formulas. We expect that the asymptotic limit of T-functions for typical representations discussed 
in subsection 3.6 is a clue for solving the problem. 
Although (ii) is not a topic of this paper, we formulated our formulas keeping in mind the construction of 
Q-operators proposed in \cite{BLZ98}. 
In particular, the parameters $\{z_{a}\}$ correspond to boundary twist parameters, and 
play a role as regulators for traces over infinite dimensional auxiliary spaces of Q-operators. 
A problem is to investigate 
how the reduction procedures discussed in this paper work on the level of Q-operators or the 
asymptotic representation theory of $U_{q}(gl(2r|1)^{(1)})$ 
\cite{BT08,T12,Z14,T19-1} (see \cite{BDKM06,KLT12,FLMS10} for the rational case). 
We will have to make a comparison with the asymptotic representation theory of $U_{q}(B_{r}^{(1)})$ \cite{HJ11,FH13}
 (see \cite{FT21} for the rational case). 
In the process, the meaning of the Boson-Fermion correspondence would be much clearer. 
In the context of (iii) (and also (ii)), it is desirable to investigate  
how the master T-operator \cite{AKLTZ11,KLT12} (generating function of transfer matrices) for $U_{q}(B_{r}^{(1)})$ 
follow from the one for $U_{q}(gl(2r|1)^{(1)})$. 
Taking note on the fact that the master T-operator is a tau-function in soliton theory (mKP for type A algebra case), 
we may end up with some kind of correspondence in soliton theory 
 (cf. KP/BKP, KdV/BKP correspondences, etc. \cite{JM83,Al20}). 
\section*{Acknowledgments} 
The work is supported by Grant No. 0657-2020-0015 of the Ministry of Science and Higher Education of Russia. 
A part of this work was done when the author was at 
Okayama Institute for Quantum Physics (where he was supported by 
Grant-in-Aid for Young Scientists, B \#19740244 from 
The Ministry of Education, Culture, Sports, Science and Technology in Japan), 
and 
 School of Mathematics and Statistics, the University of Melbourne
 (where he was supported by the Australian Research Council). 

\section*{Note added
\footnote{This note is not included in the journal version (Nuclear Physics B 972 (2021) 115563). 
There is no change in the main text. 
{\bf Comment added for arXiv:2106.08931v5:} Conditions for $\mathsf{Tab}_{I_{K}}(\lambda\subset \mu)$ in \eqref{DVF-tab1b} are corrected as below. 
We remark that 
some other algebras cases have been published in 
Nuclear Physics B 1005 (2024) 116607 [arXiv:2309.16660].} for arXiv:2106.08931v4}
\label{ApA}
\addcontentsline{toc}{section}{Appendix A}
\def\theequation{A\arabic{equation}}
\setcounter{equation}{0}
Here we summarize expressions which are necessary for precise comparison of our results and those of \cite{KOS95} (and \cite{KS94-2,KNS10}).

Based on the relation $\sum_{j \in \mathfrak{I}}p_{j}=M-N$, one can show 
\begin{align}
{\mathcal   X}_{I_{K}}^{[M-N]}&=
z_{\gamma_{K}}
\frac{\Qb_{I_{K-1}}^{[\sum_{j \in \overline{I}_{K-1}}p_{j}-2p_{\gamma_{K}}]}
\Qb_{I_{K}}^{[\sum_{j \in \overline{I}_{K}}p_{j}+2p_{\gamma_{K}}]}
}{
\Qb_{I_{K-1}}^{[\sum_{j \in \overline{I}_{K-1}}p_{j}]}
\Qb_{I_{K}}^{[\sum_{j \in \overline{I}_{K}}p_{j}]}
} ,
\label{boxes2} 
\end{align}
where $\overline{I}_{K}=(\gamma_{K+1},\gamma_{K+2},\dots,\gamma_{M+N})$. 
We rearrange \eqref{DVF-tab1} as 
${\mathcal  F}_{\lambda\subset \mu}^{I_{K}}=\check{\mathcal  F}_{\widetilde{\lambda\subset \mu}}^{I_{K}}$, 
where 
\begin{align}
\check{\mathcal  F}_{\lambda\subset \mu}^{I_{K}}=
\sum_{t \in \mathsf{Tab}_{I_{K}}(\lambda\subset \mu)}
\prod_{(j,k) \in \lambda\subset \mu}
p_{\gamma_{K+1-t_{j,k}}}
{\mathcal  X}_{I_{K+1-t_{j,k}}}^{[-\mu_{1}+\mu_{1}^{\prime}-2j+2k+m-n]}. 
\label{DVF-tab1b} 
\end{align}
Here the conditions (ii) and (iii) in \eqref{DVF-tab1} have to be modified 
for \eqref{DVF-tab1b} as:  
(ii) $t_{jk} < t_{j,k+1}$ if $\gamma_{K+1-t_{jk}}\in {\mathfrak F}$ or
  $\gamma_{K+1-t_{j,k+1}} \in {\mathfrak F}$, 
(iii)  $t_{jk} < t_{j+1,k}$ if $\gamma_{K+1-t_{jk}}\in {\mathfrak B}$ or
  $\gamma_{K+1-t_{j+1,k}} \in {\mathfrak B}$.
Under the reduction in section 3.2, 
$\check{\mathcal  F}_{\lambda\subset \mu}^{I_{M+N}}$ 
for $I_{M+N}=I_{2r+1}=(1,2,\dots, r,2r+1,r+1,r+2,\dots, 2r)$ corresponds to [eq.\ (3.2), \cite{KOS95}]. 
Here ${\mathcal   X}_{I_{2r+2-a}}^{[2r-1]}$ for $1 \le a \le r$, $a=r+1$ and $r+2 \le a \le 2r+1$ corresponds
 to 
$z(a;u)$, $z(0;u)$ and $z(\overline{2r-a+2};u)$ in [eq.\ (2.4), \cite{KOS95}], respectively.  
$I_{M+N}$ corresponds
 to $(\overline{1},\overline{2},\dots , \overline{r},0,r,\dots, 2,1)$ 
\footnote{The other option is to choose $(1,2,\dots , r,0,\overline{r},\dots, \overline{2},\overline{1})$. 
In this case,  ${\mathcal  F}_{\lambda\subset \mu}^{I_{M+N}}$ 
corresponds to $T_{\lambda\subset \mu}(-u-2r+1)$ in [eq.\ (3.2), \cite{KOS95}]. 
${\mathcal   X}_{I_{a}}$ for $1 \le a \le r$, $a=r+1$ and $r+2 \le a \le 2r+1$ corresponds
 to 
$z(a;-u)$, $z(0;-u)$ and $z(\overline{2r-a+2};-u)$ in [eq.\ (2.4), \cite{KOS95}], respectively. 
The Q-function $\Qb_{1,2,\dots ,a}$ for $a \in \{1,2,\dots r \} $ corresponds to $Q_{a}(-u)$ in \cite{KOS95}. 
Shifting  the spectral parameter $-2r+1$ in the T-functions effectively amounts to changing the sign of the unit of the shift. This 
can be seen in the boundary condition of the T-system \eqref{Tsys-bc}: 
$\Tb_{0,s}^{[-2r+1]}=\prod_{j=1}^{s}\Qb_{\emptyset}^{[-(2r-s+2j-1)]} \Qb_{\emptyset}^{[-(s-2j)]}$. 
Moreover flipping the parameters $z_{j} \to z_{j}^{-1}$ (for all $j \in \mathfrak{I}$) in the QQ-relations \eqref{QQb} and \eqref{QQf} 
  effectively amounts to changing the sign of the unit of the shift.  
  Thus the two different ways of comparison of our results and those of \cite{KOS95} are equivalent up to the convention. 
}
in the notation in \cite{KOS95}. 
The Q-function $\Qb_{1,2,\dots ,a}$ for $a \in \{1,2,\dots r \} $ corresponds to $Q_{a}(u)$ in \cite{KOS95} 
(in case the additive shift of the spectral parameter is adapted).

In particular, ${\mathcal  F}_{\mu}^{I_{K}}=\check{\mathcal  F}_{\mu}^{I_{K}}$ holds if $\mu$ is of rectangular shape. 
Then we rewrite \eqref{superJT1} as
\begin{align}
\check{\mathcal F}_{\lambda \subset \mu}^{I_{K}}&= 
\begin{vmatrix}
    \left(
    \check{\mathcal F}^{I_{K} \, [-\mu_{1}+\mu_{1}^{\prime}-\mu_{i}^{\prime}-\lambda_{j}^{\prime}+i+j-1]}_{(1^{\mu_{i}^{\prime}-\lambda_{j}^{\prime}-i+j})} \
    \right)_{1 \le i \le \mu_{1} \atop
    1 \le j \le \mu_{1}}
 \end{vmatrix}   ,
 \label{superJT1-2}
\end{align}
where the identity $|(A_{ij})_{1 \le i,j \le \mu_{1}}|=|(A_{\mu_{1}+1-j,\mu_{1}+1-i})_{1 \le i,j \le \mu_{1}}|$ for a matrix $(A_{ij})_{1 \le i,j \le \mu_{1}}$  is 
used. 
Based on \eqref{DVF-tab1b}, one can avoid the 180 degrees rotated Young diagram in \eqref{TF-rel03}. 
\begin{align}
 {\mathsf F}_{\mu}^{I_{K}}
 =
\Qb_{\emptyset}^{[m-n+\mu_{1}-\mu_{1}^{\prime}]} 
\Qb_{I_{K}}^{[-\mu_{1}+\mu_{1}^{\prime}]}
\check{\mathcal  F}_{\mu}^{I_{K}}.
\label{TF-rel03-2}
\end{align}
We also rewrite \eqref{JT-wron} and \eqref{CBRsogen} as 
$\Tb_{\lambda \subset \mu}=\check{\Tb}_{\widetilde{\lambda \subset \mu}}$, 
${\mathbb S}_{\lambda \subset \mu}=\check{\mathbb S}_{\widetilde{\lambda \subset \mu}}$, where 
\begin{align}
\check{\Tb}_{\lambda \subset \mu}&= 
\begin{vmatrix}
  \left(\Ts^{\Bm,\Fm \, [-\mu_{1}+\mu_{1}^{\prime}-\mu_{i}^{\prime}-\lambda_{j}^{\prime}+i+j-1]}_{\mu_{i}^{\prime}-\lambda_{j}^{\prime}-i+j ,1}
   \right)_{1 \le i \le \mu_{1} \atop
   1 \le j \le \mu_{1}}
  \end{vmatrix} ,
 \label{JT-wron-2}
\\
\check{\mathbb S}_{\lambda \subset \mu}&=
\begin{vmatrix}
\left(\Tb_{r,1}^{ [2i-2+2(\mu_{1}^{\prime}-\lambda_{i}^{\prime}-r)]}\right)_{1 \le i \le \mu_{1} }
&
\left(\Ts_{\mu_{j}^{\prime}-\lambda_{i}^{\prime}+i-j,1}^{\Bm, \Fm \, [2\mu_{1}^{\prime}-\mu_{j}^{\prime}-\lambda_{i}^{\prime}+i+j-r-\frac{5}{2}]}\right)_{1 \le i \le \mu_{1} \atop 2 \le j \le \mu_{1}}  
\end{vmatrix}
.
\label{CBRsogen-2}
\end{align}
One sees that 
\eqref{JT-wron-2} and \eqref{CBRsogen-2} correspond to 
(3.5a) and (4.6a) in \cite{KOS95}, respectively.  
In terms of these, \eqref{speven} and \eqref{spinodd} are expressed as
\begin{align}
\check{\Tb}_{\mu}&=\Phi_{\mu} \Ts^{\Bm,\Fm}_{\mu}
 ,
\label{speven-2}
\\
\check{\mathbb S}_{\mu +(1^{r})}^{[-\mu_{1}+\mu_{1}^{\prime}-r]}&=
\Qb_{\Bm,\Fm}^{[-\mu_{1}+\mu_{1}^{\prime}+\frac{1}{2}]} \Phi_{\mu}^{[\frac{1}{2}]} 
\prod_{j=1}^{r}(z_{j}^{\frac{1}{2}}+z_{j}^{-\frac{1}{2}})\Ts^{\Bm,\emptyset \, [-\frac{1}{2}]}_{\mu}
 .
\label{spinodd-2}
\end{align}
In particular, $\Tb_{\mu}=\check{\Tb}_{\widetilde{\mu}}$ and 
${\mathbb S}_{\mu}=\check{\mathbb S}_{\widetilde{\mu}}$ hold if $\mu$ is of rectangular shape. 
Thus the solution \eqref{CBRten}-\eqref{CBRso} has another equivalent expression:
\begin{align}
\Tb_{a,s}&=
\begin{vmatrix}
\left(\Ts_{a-i+j,1}^{\Bm, \Fm \, [-s+i+j-1]}\right)_{1 \le i \le s \atop 1 \le j \le s} 
\end{vmatrix}
\qquad \text{for} \quad 0 \le a \le r-1
.
\label{CBRten-2}
\\
\Tb_{r,2s}&=
\begin{vmatrix}
\left(\Ts_{r-i+j,1}^{\Bm, \Fm \, [-s+i+j-1]}\right)_{1 \le i \le s \atop 1 \le j \le s} 
\end{vmatrix}
,
\label{CBRse-2}
\\
\Tb_{r,2s+1}&=
\begin{vmatrix}
\left(\Tb_{r,1}^{ [-s+2i-2]}\right)_{1 \le i \le s+1 } 
&
\left(\Ts_{r+i-j,1}^{\Bm, \Fm \, [-s+i+j-\frac{5}{2}]}\right)_{1 \le i \le s+1 \atop 2 \le j \le s+1} 
\end{vmatrix}.
\label{CBRso-2}
\end{align}
In addition, \eqref{CBRso} and \eqref{CBRso-2} have two more equivalent expressions:
\begin{align}
\Tb_{r,2s+1}&=
\begin{vmatrix}
\left(\Ts_{r-i+j,1}^{\Bm, \Fm \, [-s+i+j-\frac{3}{2}]}\right)_{1 \le i \le s+1 \atop 1 \le j \le s} 
&
\left(\Tb_{r,1}^{ [-s+2i-2]}\right)_{1 \le i \le s+1 } 
\end{vmatrix}
\label{CBRso-3}
\\
&=
\begin{vmatrix}
\left(\Tb_{r,1}^{ [s-2i+2]}\right)_{1 \le i \le s+1 } 
&
\left(\Ts_{r+i-j,1}^{\Bm, \Fm \, [s-i-j+\frac{5}{2}]}\right)_{1 \le i \le s+1 \atop 2 \le j \le s+1} 
\end{vmatrix}.
\label{CBRso-4}
\end{align}
One can derive these from \eqref{CBRso} and \eqref{CBRso-2} by using \eqref{T+T} 
(cf. [page 6224 in \cite{KOS95}]). 
We remark that \eqref{CBRten-2}-\eqref{CBRso-3} correspond to [(5.1a)-(5.1c), \cite{KOS95}]. 
We also modify \eqref{Tspf} by setting $s_{i} \to -s_{i}$ ($1\le i \le r$) in the summand:
\begin{align}
\Tb_{r,1}=\sum_{s_{1},\dots , s_{r}=\pm 1}
\Qb_{\emptyset}^{[s_{1}+r-\frac{1}{2}]}
\prod_{a=1}^{r-1} 
z_{a}^{-\frac{s_{a}}{2}}
\left(
\frac{\Qb_{1,2\dots , a}^{[r-\frac{3}{2}-\frac{\rho_{a}}{2}]}}{\Qb_{1,2\dots , a}^{[r+\frac{1}{2}-\frac{\rho_{a}}{2}]}}
\right)^{\frac{s_{a}-s_{a+1}}{2}}
z_{r}^{-\frac{s_{r}}{2}}
\left(
\frac{\Qb_{1,2\dots , r}^{[r-1-\frac{\rho_{r}}{2}]}}{\Qb_{1,2\dots , r}^{[r-\frac{\rho_{r}}{2}]}}
\right)^{s_{r}}.
\label{Tspf-2}
\end{align}
In this form, one sees the same shift of the spectral parameter in the Q-functions as in [eq.(7.15) in \cite{KNS10}]. 

\end{document}